\documentclass [11pt]{article}
\usepackage{graphicx,psfrag,mathrsfs}

\catcode`@=12
\newcommand{\be}{\begin{equation}}
\newcommand{\ee}{\end{equation}}
\newcommand{\bea}{\begin{eqnarray}}
\newcommand{\eea}{\end{eqnarray}}

\def\Ds{\Delta m_{\odot}^2}
\def\Da{\Delta m_{atm}^2}
\def\s12{\sin\theta_{12}}
\def\s23{\sin\theta_{23}}
\def\s13{\sin\theta_{13}}
\def\t12{\theta_{12}}
\def\t23{\theta_{23}}
\def\t13{\theta_{13}}
\def\e{\epsilon}

\makeatletter
\@addtoreset{equation}{section}

\makeatother
\title{
{\LARGE \bf Nonzero $U_{e3}$, CP violation and leptogenesis in a see-saw type softly broken $A_4$ symmetric model}}
\author{{\bf Biswajit Adhikary$^{\rm a, b}$\footnote{biswajit.adhikary@saha.ac.in}
     and Ambar Ghosal$^{\rm a}$\footnote{ambar.ghosal@saha.ac.in}}\\
  a) Saha Institute of Nuclear Physics,\\ 1/AF Bidhan  
        Nagar, Kolkata 700064, India \\
  b)Department of Physics, Gurudas College ,\\
Narkeldanga, Kolkata-700054, India}
\date{}
\begin{document}
\maketitle
\thispagestyle{empty}
\begin{abstract}
We have shown that non-zero $U_{e3}$ is generated in a see-saw type softly broken $A_4$ symmetric model through a single parameter perturbation in $m_D$ in a single element. We have explored all possible 9 cases to study the neutrino mixing angles considering the best fitted values of $\Ds$ and $\Da$ with all parameters real. We have extended our analysis for the complex case and demonstrated  large low energy CP violation ($J_{CP}\simeq 10^{-2}$) and $m_{ee}$ in addition to mixing and mass pattern. We have also investigated leptogenesis and for a reasonable choice of model parameters compatible with low energy data, WMAP value of baryon asymmetry $6\times 10^{-10}$ is obtained for right handed neutrino mass scale $M_0\simeq 10^{13}$ GeV. We have obtained a relation among the phases responsible for leptogenesis and have shown that those phases also have correlations with low energy CP violating phases.
\end{abstract}
PACS number(s): 14.60.Pq, 11.30.Hv, 98.80.Cq
\section{Introduction}
In recent time people are too much interested to find some flavor symmetry 
in order to generate mass and mixing pattern of fermions. 
Continuous symmetry like $L_e-L_\mu-L_\tau$ \cite{lelmlt}, 
$L_\mu-L_\tau$ \cite{lmmlt} symmetry and most popular discrete 
symmetry, $\mu-\tau$ exchange symmetry ($S_2^{\mu\tau}) $\cite{mutau} 
have got some success to describe mass and mixing pattern in leptonic sector. 
To avoid mass degeneracy of $\mu$ and $\tau$ under $S_2^{\mu\tau}$ 
symmetry, E. Ma and G. Rajasekaran in \cite{Ma:2001dn} 
have introduced first time  the $A_4$ symmetry. 
After this paper, a lot of work have done with this symmetry 
\cite{Ma:2001dn}-\cite{Plentinger:2008up}. After introduction of tri-bi maximal mixing pattern ($\sin\theta_{12} = 1/\sqrt{3}$, $\sin\theta_{23} = -1/\sqrt{2}$,
$\sin\theta_{13} = 0$)\cite{tribi}, people have tried to fit this mixing pattern through the  $A_4$ symmetry. 
In an well motivated extension of the Standard model through the 
inclusion of $A_4$ discrete symmetry tri-bi maximal mixing pattern comes out in a natural way in the work of Altarelli and Feruglio \cite{Altarelli:2005yx}. More precisely, the leptonic mixing arises 
solely from the neutrino sector since the charged lepton mass matrix 
is diagonal.
The model \cite{Altarelli:2005yx} also admits  hierarchical masses 
of the three charged leptons whereas the neutrino masses 
are quasi-degenerate or hierarchical.
Although the model gives rise to
$\theta_{13} = 0$ ($U_{e3} = 0$) which is consistent with the
CHOOZ-Palo Verde experimental upper bound 
($\theta_{13}<12^\circ$ at 3$\sigma$), however, the non-zero and complex 
value of $U_{e3}$
leads to the  possibility to explore {\it CP} violation 
in the leptonic
sector 
which is the main goal of many future short and long baseline
experiments. Within the framework of $SU(2)_L\times U(1)_Y\times A_4$ model, 
non-zero $U_{e3}$ is generated either through the  
radiative correction \cite{Adhikary:2006wi} or due to the introduction of 
higher dimensional mass terms \cite{Altarelli:2005yx}. Generation of non zero complex $U_{e3}$ and possibility of non-zero CP violation has been extensively studied in \cite{Adhikary:2006jx} for the proposed model of Altarelli-Feruglio \cite{Altarelli:2005yx} with explicit soft breaking of $A_4$ symmetry \cite{Adhikary:2006wi}. 
 
In the model \cite{Altarelli:2005yx} the authors showed that the tri-bi maximal mixing pattern is also  generated naturally in the framework of see-saw mechanism with $SU(2)_L\times U(1)_Y\times A_4$ symmetry. Exact tri-bi maximal pattern forbids at low energy CP violation in leptonic sector. The textures of mass matrices in \cite{Altarelli:2005yx} could not generate lepton asymmetry also. In the present work, we investigate the generation of non-zero 
$U_{e3}$ through see saw mechanism by considering a small perturbation 
in $m_D$, the Dirac neutrino mass matrix, keeping the same texture of the 
right-handed Majorana neutrino mass matrix as proposed in Ref.\cite{Altarelli:2005yx}. 
 At first, we have studied in detail perturbation of $m_D$ by adding  
a small parameter at different entries of $m_D$ and see the variations 
of three mixing angles in terms of other model parameters considering all of them real. We extend our analysis to the complex case for a suitable texture. We study  detailed phenomenology of neutrino mass and mixing including CP violation at low energy, neutrinoless double beta decay  and leptogenesis. Our approach to get nonzero $U_{e3}$ is minimal as we break $A_4$ symmetry explicitly by single parameter in single element of $m_D$. Generation of CP violation at low energy as well as high energy is also minimal as we consider only one parameter complex. 
\section{The Model of Altarelli Feruglio with See-Saw:~light neutrino phenomenology and leptogenesis}
We consider the model proposed in \cite{Altarelli:2005yx}, which gives rise to diagonal 
$m_D$ and $M_l$ (the charged lepton mass matrix) along with a 
competent texture of $M_R$ and after see-saw mechanism and diagonalisation
gives rise to tri-bimaximal mixing pattern. The model consists of several 
scalar fields to generate required vacuum alignment to obtain tri-bimaximal 
mixing. In Table I., we have listed the scalar fields and their VEV's and representation content 
under all those symmetries. 
\begin{table}
\begin{center}
\begin{tabular}{|c|c|c|c|}
\hline
{\rm Lepton}& $SU(2)_L$ & $A_4$&\\
\hline
$\psi^{\rm l}(\nu_l, l)$&2&3&\\
$e_R$&1&1&\\
$\mu_R$&1&$1^{\prime\prime}$&\\
$\tau_R$&1&$1^{\prime}$&\\
$N_{lR}$&1&3&\\
\hline
Scalar&&&{\rm VEV}\\
\hline
$h_u$&2&1&$<h_u^0>$= $v_u/{\surd 2}$\\
$h_d$&2&1&$<h_d^0>$=$v_d/{\surd 2}$\\
$\xi$&1&1&$<\xi^0>$ = u\\
$\phi_S$&1&3&$<\phi_S^0>$ = $(v_S,v_S,v_S)$\\
$\phi_T$&1&3&$<\phi_T>$ = $(v_T,0,0)$\\
\hline
\end{tabular}
\end{center}
\caption{List of fermion and scalar fields used in this  model, $l=e,~\mu,~\tau$.
}
\end{table}
 The 
model is fabricated in such a way that after spontaneous breaking of $A_4$ symmetry, the $S_2^{\mu\tau}$ symmetry remains 
on the neutrino sector and the charged lepton sector is invariant 
under $Z_3$ symmetry. 
Consider the Lagrangian of the model \cite{Altarelli:2005yx},
\begin{eqnarray}
{\mathcal L} &&= \frac{y_e}{\Lambda}(\phi_T {\bar \psi}^{\rm l}_L)e_Rh_d + \frac{y_\mu}{\Lambda} (\phi_T {\bar \psi}^{\rm l}_L)^\prime \mu_Rh_d
         + \frac{y_\tau}{\Lambda} (\phi_T {\bar \psi}^{\rm l}_L)^{\prime\prime}\tau_R h_d +f{\bar \psi}^{\rm l}_L N_R h_u \nonumber\\
&&+ x_A \xi{\bar N}^c_LN_R + x_B \phi_S{\bar N}^c_LN_R+ h.c. 
 \label{lag} 
\end{eqnarray} 
After spontaneous symmetry breaking, the charged lepton mass matrix 
comes out diagonal with $m_e=\frac{y_ev_Tv_d}{\sqrt{2}\Lambda}$, $m_\mu=\frac{y_\mu v_Tv_d}{\sqrt{2}\Lambda}$, and $m_\tau=\frac{y_\tau v_Tv_d}{\sqrt{2}\Lambda}$. The neutrino sector gives rise to the 
following Dirac and Majorana matrices 
\begin{eqnarray}
m_D = f \frac{v_u}{\surd 2}\pmatrix{1&0&0\cr
                    0&1&0\cr
                    0&0&1}\qquad
M_R=\pmatrix{A + 2D/3 & -D/3 & -D/3\cr
                     -D/3     & 2D/3& A - D/3\cr
                     -D/3     & A - D/3 & 2D/3}
 \label{mdmr} 
\end{eqnarray} 
where $A=2x_Au$, $D=2x_B v_S$. The structure of light neutrino mass matrix can be obtained from see-saw formula:
\begin{eqnarray}
M_\nu=-m_DM_R^{-1}m_D^T=U_{TB} \pmatrix{\frac{-f^2v_u^2}{2(D+A)}&0&0\cr
                    0&\frac{-f^2v_u^2}{2A}&0\cr
                    0&0&\frac{-f^2v_u^2}{2(D-A)}}U_{TB}^T
\label{ssf} 
\end{eqnarray} 
where,
\begin{eqnarray}
U_{TB}= \pmatrix{\sqrt{\frac{2}{3}}& \sqrt{\frac{1}{3}} &0\cr
                    -\sqrt{\frac{1}{6}}& \sqrt{\frac{1}{3}} & -\sqrt{\frac{1}{2}}\cr
                    -\sqrt{\frac{1}{6}}&\sqrt{\frac{1}{3}}& \sqrt{\frac{1}{2}}}.
\label{tbmix} 
\end{eqnarray} 
This is clear from Eq.\ref{ssf} that $U_{TB}$ is the diagonalising matrix for light neutrino mass matrix $M_\nu$. The form of $U_{TB}$ is in Eq.\ref{tbmix} which is nothing but the so called tribimaximal mixing matrix. From Eq.\ref{ssf} we have the eigenvalues of $M_\nu$:
\begin{eqnarray}
 m_1=-\frac{f^2v_u^2}{2(D+A)}\qquad m_2=-\frac{f^2v_u^2}{2A}\qquad m_3=-\frac{f^2v_u^2}{2(D-A)}
\label{a4ev} 
\end{eqnarray} 
 From Eq.\ref{tbmix} we have the mixing angles $\sin\theta_{12} = 1/\sqrt{3}$, $\sin\theta_{23}=-1/\sqrt{2}$ 
 and $\sin\theta_{13} = 0$ and from Eq.\ref{a4ev} we get the solar and atmospheric mass squared differences as
\begin{eqnarray}
&&\Ds =m_2^2-m_1^2= \frac{m_0^2k(k+2)}{(1+k)^2}\nonumber\\
&&\Da = m_3^2-m_2^2=\frac{m_0^2k(2-k)}{(1-k)^2}
\label{a4msd} 
\end{eqnarray} 
where $D=kA$, $m_0=\frac{f^2v_u^2}{2A}$  and all parameters are real. From the experiments we know $\Ds$ is positive and dictates either  $k>0$ or  $k<-2$. If $k>0$, then it has to be small in order to generate small value of $\Ds$ provided $m_0^2$ is not too small as $\Ds$. But small positive $k$ corresponds to same order of magnitude of $\Ds$ and $\Da$ which is not acceptable according to the experimental results. Now $k>0$ only acceptable for $m_0^2\simeq\Ds$ and hierarchy of $\Ds$ and $\Da$ obtained with the singular nature of $\Da$ as in Eq.\ref{a4msd} near $k\simeq 1$. This corresponds to normal hierarchical mass spectrum. Again for $m_0^2\gg\Ds$, $k<-2$ is the physical region. This region of $k$ makes  $\Da<0$ which is so called inverted ordering of neutrino mass pattern. Again $k+2$ should take small value in order to generate small value of $\Ds$. For one complex parameter $D\equiv De^{i\phi}$, we can write the mass differences in the following form
\begin{eqnarray}
&&\Ds = \frac{m_0^2k(k+2\cos\phi)}{1+k^2+2k\cos\phi}\nonumber\\
&&\Da = \frac{m_0^2k(2\cos\phi-k)}{1+k^2-2k\cos\phi}.
\label{a4msdc} 
\end{eqnarray} 
In the complex case, positivity of $\Ds$ can be obtained either with $k>0$ and $\cos\phi>-k/2$ or with $k<0$ and $\cos\phi<-k/2$. For the first case  with $m_0^2\simeq\Ds$ and with $\cos\phi\simeq(1+k^2)/2k$ one can have normal hierarchical mass spectrum. For the second case hierarchy will be inverted and $k+2\cos\phi$ have to be small. In both case $k$ should take the value such that the $1\ge\cos\phi\ge -1$ range also satisfy. The mixing pattern is tri-bi maximal Eq.\ref{tbmix} and it is independent to the fact whether the parameters are real or complex. In this mixing pattern $U_{e3}=0$ and non-zero complex $U_{e3}$ is  a basic requirement to see the non-zero Dirac CP violation. 

Now we concentrate on the issue of leptogenesis in this model. The decay of right handed heavy Majorana neutrinos to lepton(charged or neutral) and scalar(charged or neutral) generate non-zero lepton asymmetry if i) C and CP are violated, ii)lepton number is violated and iii) decay of right handed neutrinos are out of equilibrium. We are in the energy scale where $A_4$ symmetry is broken but the SM gauge group remains unbroken. So, the higgs scalars both charged and neutral are physical.
The CP asymmetry of decay is characterized by a parameter $\varepsilon_i$
which is defined as
\begin{eqnarray}
\varepsilon_i&=&\frac{\Gamma_{{N}_i\rightarrow
    l^-\phi^+,\nu_l\phi^0}-\Gamma_{{N}_i\rightarrow
    l^+\phi^-,\nu_l^c\phi^{0*}}}{\Gamma_{{N}_i\rightarrow
    l^-\phi^+,\nu_l\phi^0}+\Gamma_{{N}_i\rightarrow
    l^+\phi^-,\nu_l^c\phi^{0*}}}.
\label{cpasym}
\end{eqnarray}
Spontaneous $A_4$ symmetry breaking generates right handed neutrino mass and the mass matrix $M_R$ obtained is shown in Eq. \ref{mdmr}. We need to diagonalize $M_R$ in order to go into the physical basis (mass basis) of right handed neutrino. This form of $M_R$ gives the diagonalising matrix in the tri-bi maximal form $U_{TB}$ in Eq.\ref{tbmix}:
\begin{eqnarray}
U_{TB}^{\dagger}M_RU_{TB}^{*}={\rm diag}(M_1,~ M_2,~ M_3)={\rm diag}(A+D,~ A,~ D-A),
\label{a4mrd}  
\end{eqnarray} 
however, the  eigenvalues are not real. We need to multiply one diagonal  phase matrix $U_P$ with $U_{TB}$. Hence, diagonalising matrix $V=U_{TB}U_P$ relates the flavor basis to eigen basis of right handed
neutrino:
\begin{eqnarray}
N_{lR}=\sum_{i=1}^{3}V_{li}^*N_{iR}.
\label{fmr}
\end{eqnarray}
In this basis the couplings of $N_R$ with leptons and scalars are modified and it will be:
\begin{eqnarray}
m_d'= m_dV^{*}.
\label{mmd} 
\end{eqnarray} 
At the tree level there there are no asymmetry in the decay of right handed neutrinos. Due to the interference between tree level and one loop level diagrams, the asymmetry is generated. There are vertex diagram and self energy diagram to contribute to the asymmetry \cite{Fukugita:1986hr,leptogen}. The vertex contribution is :
\begin{eqnarray}
\varepsilon_i^V=\frac{1}{4\pi v_u^2h_{ii}}\sum_{j\ne
  i}Im(h_{ij}^2)\times \left[\sqrt{x_{ij}}\left\{1-(1+x_{ij})\ln(1+\frac{1}{x_{ij}})\right\}\right]
\label{vertex}
\end{eqnarray}
and the self energy part is :
\begin{eqnarray}
\varepsilon_i^S=\frac{1}{4\pi v_u^2h_{ii}}\sum_{j\ne
  i}Im(h_{ij}^2)\times\frac{\sqrt{x_{ij}}}{1-x_{ij}}.
\label{self}
\end{eqnarray}
where $x_{ij}=M_j^2/M_i^2$ and
\begin{eqnarray}
h=m_D'^\dagger m_D'.
\label{h}
\end{eqnarray}
The key matrix, whose elements are necessary to calculate leptogenesis, is $h$. In this model $m_D$ is diagonal and proportional to identity. Hence, $h$ matrix is real diagonal and it is also proportional to identity matrix and it is independent of the form of $V$. The terms for decay asymmetry generated  by  ``i'' th generation of right handed neutrino $N_{Ri}$ for both vertex and self energy contributions are proportional to ${\rm Im}(h_{ij}^2)$ (where $j\ne i$) as in Eq.\ref{vertex} and Eq. \ref{self}. All off-diagonal elements of $h$ are zero. So, decay of all three generation of right handed Majorana neutrinos could not generate lepton asymmetry. So, in this model of $A_4$ symmetry tri-bi maximal mixing pattern is not compatible with the low energy Dirac CP violation as well as high energy CP violation. In order to obtain non-zero $\t13$, low energy Dirac CP violation and leptogenesis we need to break the $A_4$ symmetry through not only spontaneously but also explicitly introducing some soft $A_4$ symmetry breaking (soft in the sense that the breaking parameter is small to consider $A_4$ as an approximate symmetry) terms in the Lagrangian.
\section{Explicit $A_4$ symmetry breaking and real parameter analysis}
\label{sec:a4brk} 
We consider minimal breaking of $A_4$ symmetry through a single parameter in a single element of $m_D$ keeping $M_R$ unaltered as 
\begin{eqnarray}
 {\mathcal L_{A_4\rm breaking}}=f\e{\bar \psi}^{\rm l}_{\alpha L} N_{\beta R} h_u.
\label{la4b} 
\end{eqnarray} 
We introduce the breaking by small dimensionless parameter $\e$ to the $(\alpha,\beta)$ element of Dirac type Yukawa term for neutrino. After spontaneous $SU(2)_L\times U(1)_Y$ symmetry breaking it modifies only one element $(\alpha,\beta)$ of $m_D$ of neutrino. There are nine possibilities  to incorporate the breaking parameter $\e$ in $m_D$. We know that after spontaneous $A_4$ symmetry breaking, a residual $S_2^{\mu\tau}$ symmetry appears in neutrino sector. There is a special feature of  $S_2^{\mu\tau}$ symmetry which ensures one $0$ and one maximal $\pi/4$ mixing angles. There is one task to check whether our newly introduced explicit breaking term can break  $S_2^{\mu\tau}$ symmetry or not. This is important because we need non-zero $\t13$. We have seen that in one case out of the nine possibilities, residual $S_2^{\mu\tau}$ symmetry remains invariant. This is $\alpha\beta=11$ case. In other cases  $S_2^{\mu\tau}$ symmetry is broken and one expect non-zero $\t13$ from those cases. Primarily, we consider that all parameters are real. We want to study the mixing pattern and want to see its deviation from tri-bi maximal pattern considering experimental value of mass squared  differences of neutrinos. We explore all nine cases including $\alpha\beta=11$ case. Although $\alpha\beta=11$ case could not generate non-zero $\t13$, however, we want to see whether this breaking can reduce the tri-bi maximal value of $\theta_{12}$ ($\simeq 35.26^\circ$) to its best fit value ($\simeq 34^\circ$) or not along with the special feature $\t13=0$ and $\theta_{23}=\pi/4$. Here, we explicitly demonstrate the procedure for a single  case and for the other cases expressions for eigenvalues and mixing angles are given in Apendix.

(i) Breaking at '22' element : In this case, the structure of 
$m_D$ is given by 
\begin{eqnarray}
m_D = \frac{fv_u}{\surd 2}\pmatrix{1&0&0\cr
                      0&1+\e&0\cr
                      0&0&1}
 \label{bmd22} 
\end{eqnarray} 
and after implementation of see-saw mechanism keeping the same 
texture of $M_R$, three light neutrino mass eigenvalues 
come out as 
\begin{eqnarray}
m_1 = -\frac{f^2v_u^2}{2(D+A)}\left(1+\frac{\e}{3}\right)
\quad\quad 
m_2 = -{\frac{f^2v_u^2}{2A}}\left(1 + \frac{2\e}{3}\right)
\quad\quad
m_3 = -\frac{f^2v_u^2}{2(D-A)}\left(1 +\e\right)
\end{eqnarray} 
and the three mixing angles come out as 
\begin{eqnarray}
\sin\theta_{12} = {\frac{1}{\sqrt 3}} - {\frac{\e(2A+D)}{3{\sqrt 3}D}}&&
\quad\quad
\sin\theta_{23} = -\left[{\frac{1}{\sqrt 2}} + {\frac{\epsilon}{3}}
\left({\frac{D}{2{\sqrt 2}A}} +
{\frac{{\sqrt 2}D}{4A - 2D}}\right)\right]\nonumber\\
\sin\theta_{13} &=& {\frac{\epsilon}{3}}
\left({\frac{D}{{\sqrt 2}A}} -{\frac{{\sqrt 2}D}{4A - 2D}}\right)
\label{angel22} 
\end{eqnarray} 
Assuming a relationship between the parameters $D$ and $A$ as 
$D = k A$
we rewrite in a convenient way the above three mixing angles as 
\begin{eqnarray}
\sin\theta_{12} = {\frac{1}{\sqrt 3}} - 
{\frac{\e(2+k)}{3{\sqrt 3}k}} \qquad\sin\theta_{23} = -\left[{\frac{1}{\sqrt 2}} + 
{\frac{\e k(4-k)}{6\sqrt 2(2-k)}}\right]
\qquad\sin\theta_{13} = \frac{\e k(1-k)}{3{\sqrt 2}(2-k)}
\label{mangel22} 
\end{eqnarray} 
and the mass-squared differences are 
\begin{eqnarray}
\Ds = \frac{m_0^2}{3(1+k)^2}\left[
3k(k+2)+2\e(2k^2+4k+1)\right]\nonumber\\ \Da = \frac{m_0^2}{3(1-k)^2}\left[
3k(2-k)-2\e(2k^2-4k-1)\right]
\label{masd22}
\end{eqnarray} 
where $m_0 = f^2v_u^2/2A$. Defining the ratio $R$ in terms of mass-squared 
differences we get
\begin{eqnarray}
 R = \frac{\Ds}{\Da} 
  = \frac{(k -1)^2}{(k+1)^2}
    \frac{\left[3k(k+2)+2\e(2k^2+4k+1)\right]}
    {\left[3k(2-k)-2\e(2k^2-4k-1)\right]}
\label{R22} 
\end{eqnarray} 
which in turn determines the parameter $\e$ as 
\begin{eqnarray}
 \e = \frac{3k}{2}
     \frac{\left[
    R(2-k)(k+1)^2-(k-1)^2(k+2)\right]}
    {\left[(k-1)^2(2k^2+4k+1)+
    R(k+1)^2(2k^2-4k-1)\right]}
\label{ep22} 
\end{eqnarray} 
Similarly, we have evaluated all other possible cases which we have 
listed in the Appendix.

\begin{figure}
\begin{center}
\includegraphics[height=8cm,keepaspectratio]{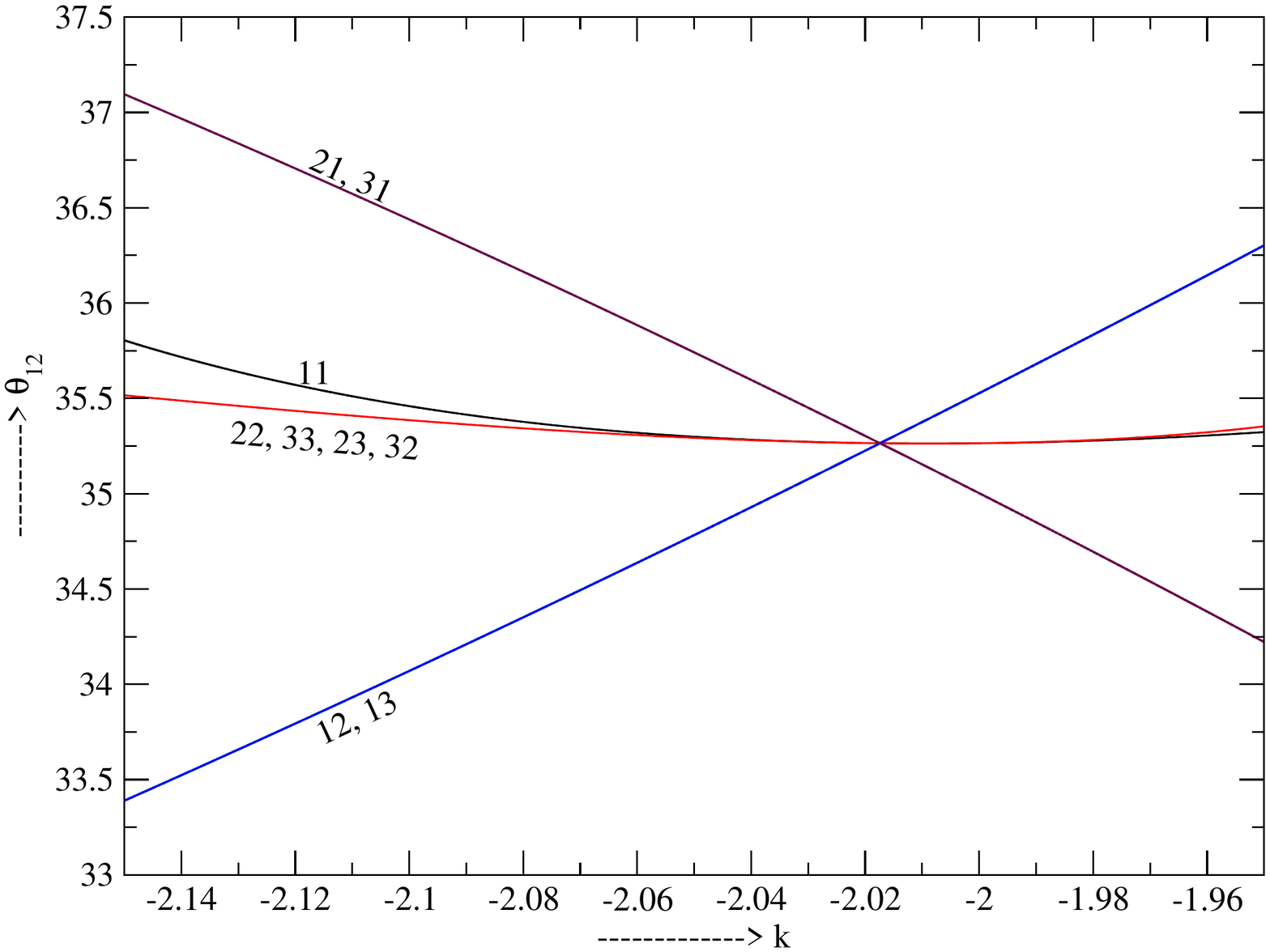}
\caption{\label{th12} Plot of   $\theta_{12}$  with respect to
  $k$ . We keep
  $\Delta m^2_{32}$ and  $\Delta m^2_{21}$ to their
  best fit values.  }
\end{center}
\end{figure}
%
\begin{figure}
\begin{center}
\includegraphics[height=8cm,keepaspectratio]{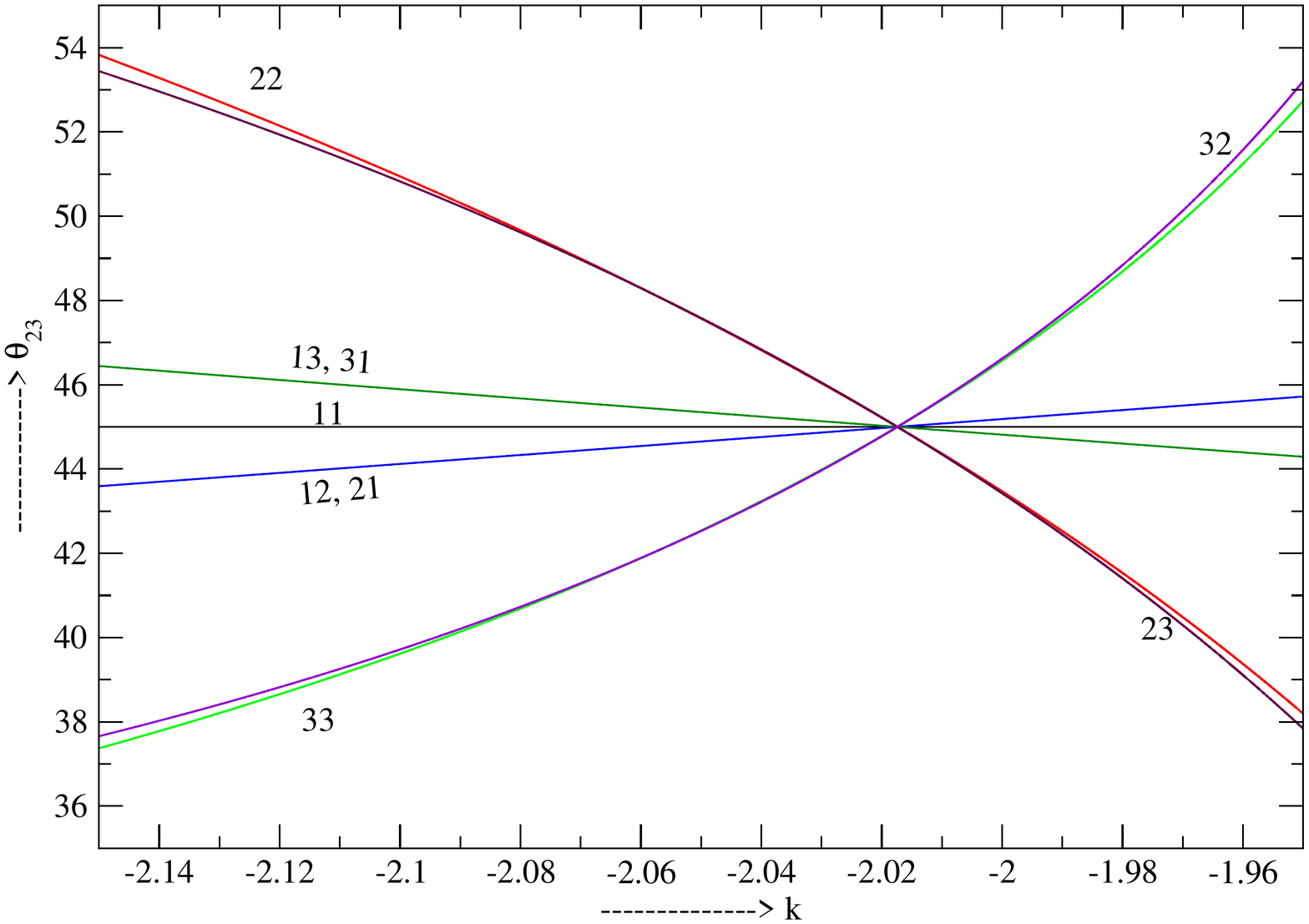}
\caption{\label{th23} Plot of   $\theta_{23}$  with respect to
  $k$. We keep
  $\Delta m^2_{32}$ and  $\Delta m^2_{21}$ to their
  best fit values.  }
\end{center}
\end{figure}
%
\begin{figure}
\begin{center}
\includegraphics[height=8cm,keepaspectratio]{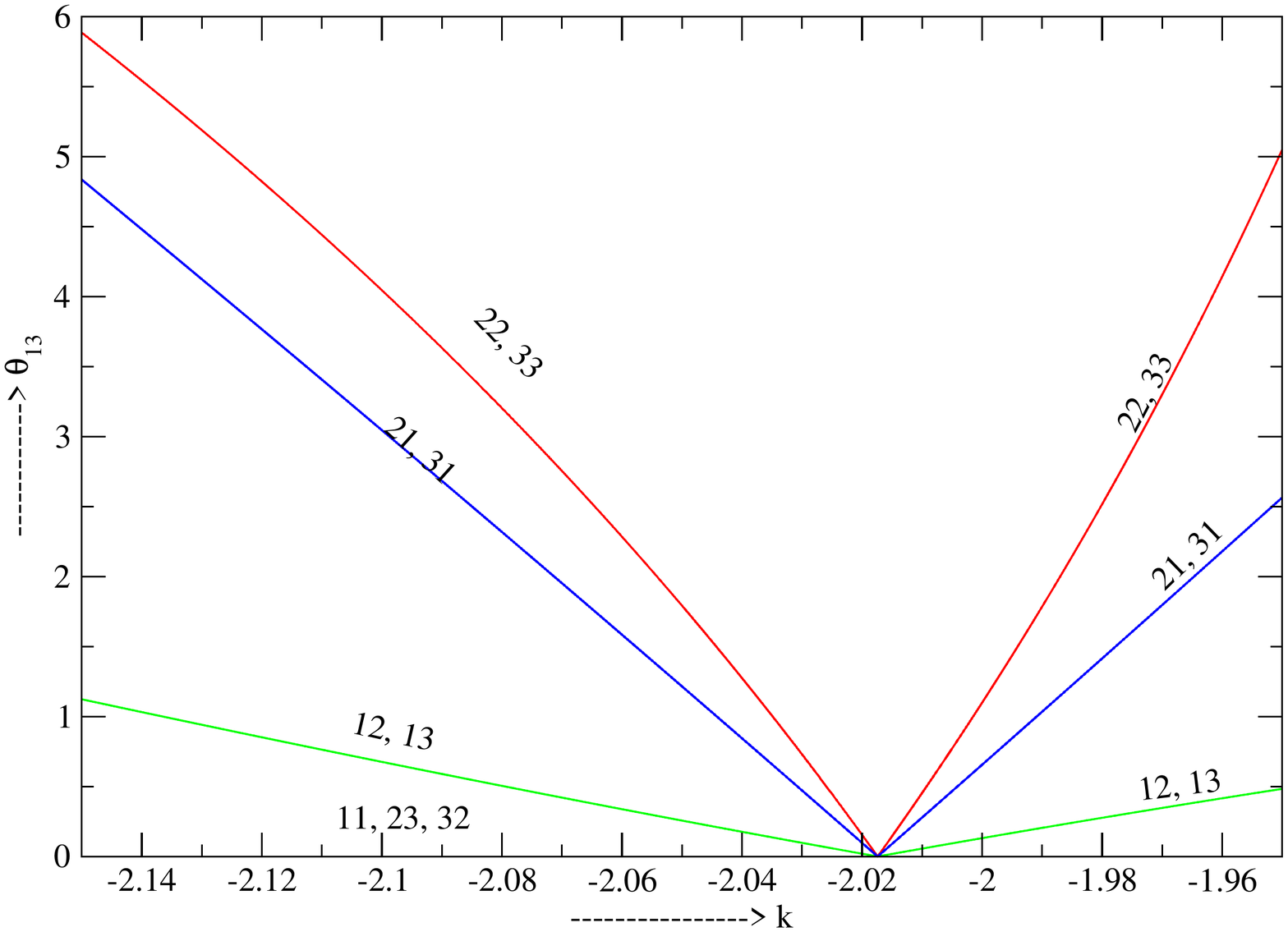}
\caption{\label{th13} Plot of   $\theta_{13}$  with respect to
  $k$. We keep
  $\Delta m^2_{32}$ and  $\Delta m^2_{21}$ to their
  best fit values.  }
\end{center}
\end{figure}
%
Now the $\epsilon$ parameter is determined in terms of $R$ and $k$ 
and we substitute it to the expressions for mixing angles. Thus it is possible to explore 
all three mixing angles $\theta_{12}$, $\theta_{23}$ and $\theta_{13}$
in terms of $R$ and $k$. Particularly, the deviation from 
tri-bimaximal mixing depends only on $R$ and $k$. For the best fit values of the solar and atmospheric mass squared differences  
($R\simeq 4\times{10}^{-2}$),  we have shown in
Fig. \ref{th12}, Fig. \ref{th23} and Fig. \ref{th13}  the variations of $\theta_{12}$, $\theta_{23}$ 
, $\theta_{13}$ verses $k$, respectively. We have studied all 
nine possible cases and shown in the plots. 
First of all, non-zero value of $\theta_{13}$ is obtained if we allow 
$A_4$ symmetry breaking terms explicitly  
in any one of the '12, 13, 21, 31, 22, 33' element of the Dirac neutrino 
mass matrix and those are the cases of our interest.
In the present analysis, we have shown that  non-zero $\theta_{13}$ is generated in a 
softly broken $A_4$ symmetric model which leads to deviation from the 
tri-bimaximal mixing. In $A_4$ symmetric model
$\theta_{13}$ is zero because of residual $S_2^{\mu\tau}$ symmetry in neutrino sector after spontaneous 
breaking of $A_4$ symmetry. Apart from the explicit breaking of $A_4$ symmetry at 11 element,  $S_2^{\mu\tau}$ is broken for all '12, 13, 21, 31, 22, 33, 23, 32' cases. Furthermore, perturbation around 
'23', '32' elements also lead to zero value of $\theta_{13}$ at the 
leading order although $S_2^{\mu\tau}$ symmetry is broken, non-zero value 
is generated if we consider higher order terms of $\epsilon^2$ which are 
too tiny and hence, discarded from our analysis.  We include 11 case for completeness which preserves $S_2^{\mu\tau}$ symmetry and hence generates $\theta_{13}=0$ and $\theta_{23}=\pi/4$. It only shifts $\theta_{12}$ from the tri bimaximal value, but it cannot be able to go towards the best fit value of solar angle , $34^\circ$.

If $k = -2$, then we get $\epsilon\propto R$, and thereby,
the value of $\theta_{13}$ is very small also $\theta_{12}$ will 
hit the exact tri-bimaximal value in some cases. The effect of variation on the mixing angles around $k=-2$  are asymmetric. For some cases (for example 23, 32) $\theta_{23}$ changes very fast in the $k>-2$ region. So, we explore the mixing angles with the range $-2.15\le k \le -1.95$. We choose the most feasible cases in which perturbation 
is applied around '12', '13' elements, because in those cases, 
variation of $k$ encompasses the best-fit values of 
$\theta_{12}$ and $\theta_{23}$. Although, in the '21', '31' cases, 
the value of $\theta_{23}$ touches the best-fit value $\pi/4$, 
however, $\theta_{12}$ far apart from the best-fit value. 
In order to achieve large $\theta_{13}$, we have to choose 
the '21', '31' cases, but we have to allow the variation of 
$\theta_{12}$ around as large as $ 37.2^\circ$.
In case of '22', '33', the structure of $m_D$ is still 
diagonal and also we can get larger $\theta_{13}$(upto $6^\circ$) and also 
$\theta_{12}$ is within $1\sigma(36^\circ)$, however, $\theta_{23}$ 
will reach $3\sigma (54^\circ)$ value.

In summary, we have shown that  non-zero $\theta_{13}$ is generated in a 
$A_4$ symmetric model which leads to deviation from the 
'tri-bimaximal' mixing through see-saw mechanism due to the 
incorporation of an explicit $A_4$ symmetry breaking term 
in $m_D$. The breaking is incorporated through a single parameter 
$\epsilon$ and we have investigated the effect of such breaking term 
in all nine elements of $m_D$. Some of them generates still zero 
value of $\theta_{13}$ and rest of the others generated non-zero 
$\theta_{13}$. We expressed all three mixing angles in terms 
of one model parameter and showed the variation of all three 
mixing angles with the model parameter $k$. We find 
breaking through '12' and '13' elements of $m_D$ are most 
feasible in view of recent neutrino experimental results. 
\section{Complex extension: Light neutrino phenomenology and Leptogenesis}
\label{sec:cmlpx}
In this section, we consider one of the parameter is complex and out of all nine cases as mentioned earlier, we investigate one suitable case arises due to '13' element perturbation. This is one of the suitable positions of breaking justified from real analysis. Again this extension is minimal to generate non-zero CP violation because we  consider only one parameter complex. We take $D$ as complex: $D\equiv De^{i\phi}$. Hence, the form of $m_D$ and $M_R$  under explicit $A_4$ symmetry breaking with complex extension are :
\begin{eqnarray}
m_D = f \frac{v_u}{\surd 2}\pmatrix{1&0&\e\cr
                    0&1&0\cr
                    0&0&1}\qquad
M_R=\pmatrix{A + (2De^{i\phi})/3 & -(De^{i\phi})/3 & -(De^{i\phi})/3\cr
                     -(De^{i\phi})/3     & (2De^{i\phi})/3& A - (De^{i\phi})/3\cr
                     -(De^{i\phi})/3     & A - (De^{i\phi})/3 & (2De^{i\phi})/3}. \nonumber\\
\label{mrmdcm13} 
\end{eqnarray}
\subsection{Light neutrino phenomenology}
 Using the see-saw mechanism we get the light neutrino mass matrix as
\begin{eqnarray}
&&M_\nu = -m_DM_R^{-1}md^T=\frac{-f^2v_u^2}{2}\times\\&&U_{TB} \pmatrix{\frac{1}{De^{i\phi}+A}-\frac{2\e}{3(De^{i\phi}+A)} -\frac{2De^{i\phi}(2A+De^{i\phi})\e^2}{9A(A^2-D^2e^{i2\phi})}& \frac{(A+2De^{i\phi})\e}{3\sqrt{2}(De^{i\phi}+A)}- \frac{\sqrt{2}De^{i\phi}(2A+De^{i\phi})\e^2}{9A(A^2-D^2e^{i2\phi})}&\frac{-\e}{\sqrt{3}(A-De^{i\phi})}\cr
\frac{(A+2De^{i\phi})\e}{3\sqrt{2}(De^{i\phi}+A)} -\frac{\sqrt{2}De^{i\phi}(2A+De^{i\phi})\e^2}{9A(A^2-D^2e^{i2\phi})}&\frac{1}{A}+\frac{2\e}{3A} -\frac{De^{i\phi}(2A+De^{i\phi})\e^2}{9A(A^2-D^2e^{i2\phi})}&\frac{-\e}{\sqrt{6}(A-De^{i\phi})}\cr
\frac{-\e}{\sqrt{3}(A-De^{i\phi})}&\frac{-\e}{\sqrt{6}(A-De^{i\phi})}&\frac{1}{De^{i\phi}-A}}U_{TB}^T.\nonumber 
\label{mnuc} 
\end{eqnarray} 
We need to diagonalize the $M_\nu$ to obtain the masses and mixing angles. The eigenvalues are same as we have in the real case and only difference is that the $D$ is complex now. We explicitly write down the complex phase in the mass matrix. The obtained eigenvalues are :
\begin{eqnarray}
m_1 = -\frac{f^2v_u^2}{2(De^{i\phi}+A)}\left(1-\frac{2\e}{3}\right)
\quad\quad 
m_2 = -\frac{f^2v_u^2}{2A}\left(1+\frac{2\e}{3}\right)
\quad\quad 
m_3 = -\frac{f^2v_u^2}{2(De^{i\phi}-A)}
\label{msevc} 
\end{eqnarray} 
where we keep terms upto first  order in $\e$. Now with  $D=kA$, $m_0=f^2v_u^2/2A$ and keeping term upto first order in $\e$ we get the three neutrino mass squared as
\begin{eqnarray}
 \left|m_1\right|^2=\frac{m_0^2(9-12\e)}{9(1+k^2+2k\cos\phi)}\qquad\left|m_2\right|^2=\frac{m_0^2(9+12\e)}{9}\qquad
\left|m_3\right|^2=\frac{m_0^2}{1+k^2-2k\cos\phi}.\nonumber\\
\label{msq} 
\end{eqnarray} 
Using those expressions we get the mass squared differences and their ratio which are,
\begin{eqnarray}
\Ds=\left|m_2\right|^2-\left|m_1\right|^2=\frac{m_0^2\{9k(k+2\cos\phi)+12\e(2+k^2+2k\cos\phi)\}}{9(1+k^2+2k\cos\phi)}\nonumber\\
\Da=\left|m_3\right|^2-\left|m_2\right|^2=\frac{m_0^2\{9k(2\cos\phi-k)-12\e(1+k^2-2k\cos\phi)\}}{9(1+k^2-2k\cos\phi)}
\label{msdc13} 
\end{eqnarray} 
and
\begin{eqnarray}
 R=\frac{\Ds}{\Da}=\frac{1+k^2-2k\cos\phi}{1+k^2+2k\cos\phi}\left[\frac{9k(k+2\cos\phi)+12\e(2+k^2+2k\cos\phi)}{9k(2\cos\phi-k)-12\e(1+k^2-2k\cos\phi)}\right].
\label{rc13} 
\end{eqnarray} 
The mixing angles are obtained from diagonalisation of $M_\nu$. We solve the equations of the form $M_\nu \left|\right . m_i\left. \right>^*=m_i\left|\right . m_i\left. \right>$. These $\left|\right . m_i\left. \right>$ will give  the columns of the diagonalising unitary matrix $U$. Throughout our calculation we assume that breaking parameter $\e$ is small. We have the nonzero $U_{13}$ which is proportional to $\e$. So, the values of $U_{12}$ and $U_{23}$ will give the solar and atmospheric mixing angles, respectively. The expressions for the mixing angles come out as 
\begin{eqnarray}
\sin\theta_{12}=\left|U_{12}\right|=\frac{1}{\sqrt{3}}+\frac{\e}{3\sqrt{3}}\times\frac{2+4k^2+2k^4+9k\cos\phi+10k^2\cos^2\phi+9k^3\cos\phi}{k(1+k^2+2k\cos\phi)(k+2\cos\phi)}
\label{t12c13} 
\end{eqnarray} 
\begin{figure}
\begin{center}
\includegraphics[height=8cm,keepaspectratio]{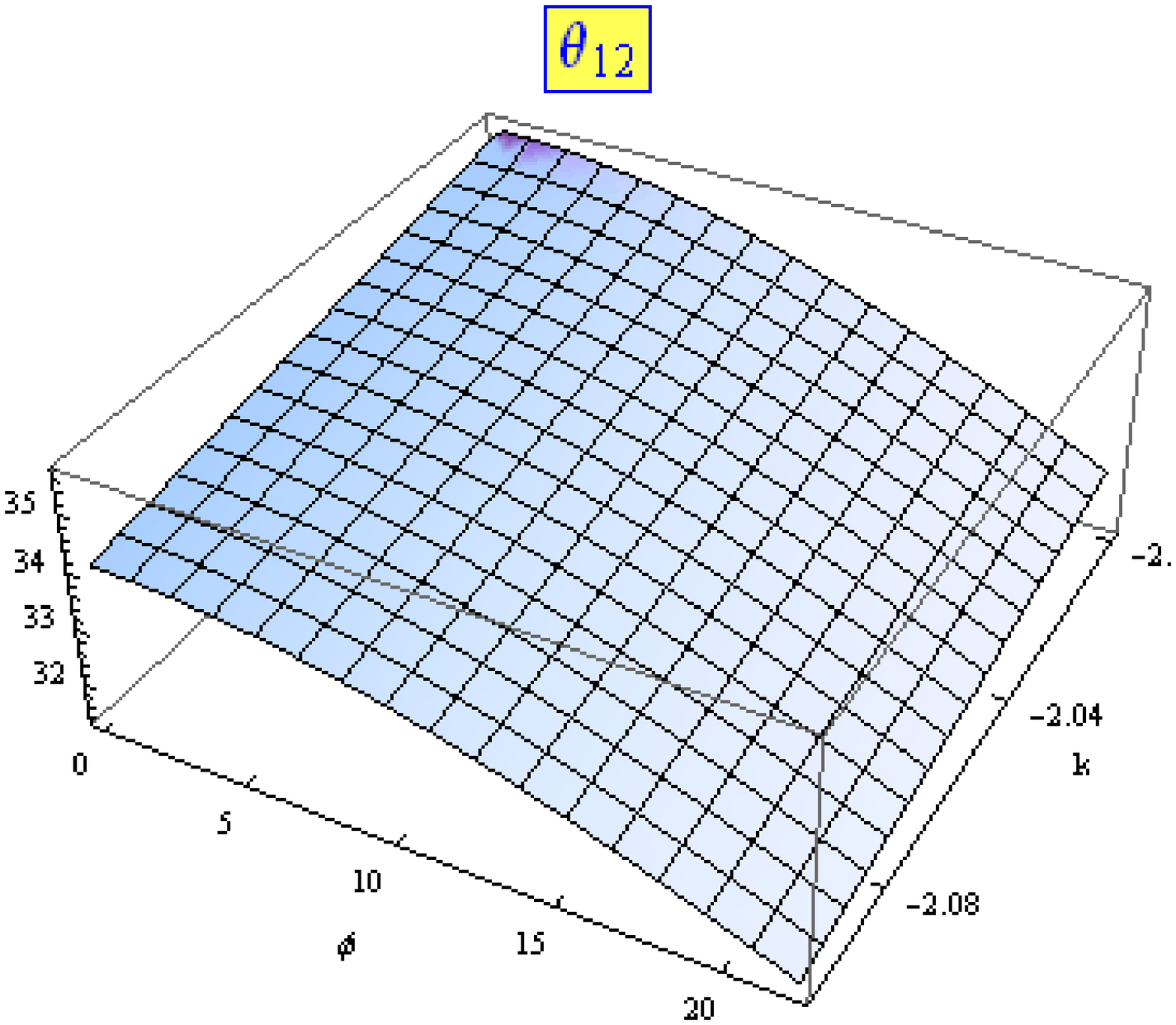}
\caption{\label{th12c13} Plot of $\theta_{12}$  with respect to
  $k$ and  $\phi$ for the breaking of $A_4$ in '13' element of $m_D$. We keep
  $\Delta m^2_{32}$ and  $\Delta m^2_{21}$ to their
  best fit values.}
\end{center}
\end{figure}
%
\begin{eqnarray}
 \sin\theta_{23}=\left|U_{23}\right|=\frac{1}{\sqrt{2}}+\frac{\e k}{6\sqrt{2}\cos\phi}\times\frac{k^3+k+2k\cos^2\phi-3k^2\cos\phi-\cos\phi}{k^3+k+4k\cos^2\phi-4k^2\cos\phi-2\cos\phi}
\label{t23c13} 
\end{eqnarray} 
\begin{eqnarray}
&&\sin\theta_{13}= \left|U_{13}\right|=\frac{\e}{3\sqrt{2}\cos\phi(k^3+k+4k\cos^2\phi-4k^2\cos\phi-2\cos\phi)}\times\nonumber\\
&&\left[(k^4+k^2-3k^3\cos\phi-k\cos\phi+6k\cos^3\phi-3\cos^2\phi-k^2\cos^2\phi)^2\right . \nonumber\\&&+\left .\sin^2\phi(k^3+k+6k\cos^2\phi-5k^2\cos\phi-3\cos\phi)^2\right]^{1/2}
\label{t13c13}. 
\end{eqnarray} 
From the expression of mixing angles it is clear that the deviations from tri-bi maximal are first order in $\e$. The  independent parameters in this model are $A$, $D$, $f$, $\e$ and $\phi$. Alternatively the independent parameters are $M_0$, $m_0$, $k$, $\e$ and $\phi$ (where $M_0=A$,$m_0=f^2v_u^2/2A$, $k=D/A$). In the above analysis of light neutrino mass and mixing, scale $M_0$ did not appear explicitly. We have four well measured  observable which are $\Ds$, $\Da$, $\theta_{12}$ and $\theta_{23}$, and, thus, in principle it is possible to determine four parameters $m_0$, $k$, $\e$ and $\phi$ and we are able to predict the other less known observable such as angle $\theta_{13}$, CP violating parameter $J_{CP}$ etc. It is difficult to get inverse relations of those observable. From the expression of $R$ in Eq. \ref{rc13}
we easily obtain the expression for $\e$ as
\begin{figure}
\begin{center}
\includegraphics[height=8cm,keepaspectratio]{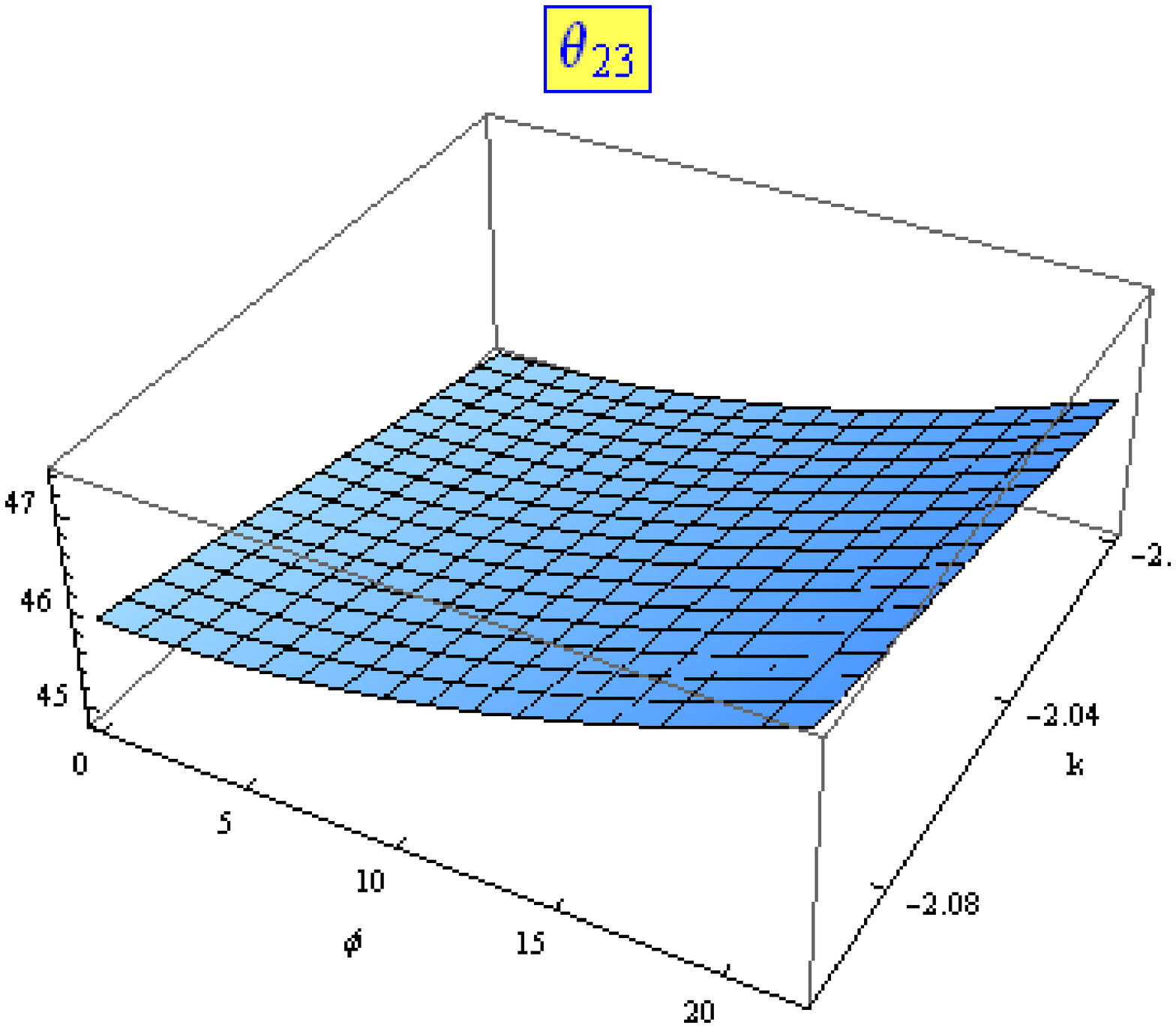}
\caption{\label{th23c13} Plot of   $\theta_{23}$  with respect to
  $k$ and  $\phi$ for the breaking of $A_4$ in '13' element of $m_D$. We keep
  $\Delta m^2_{32}$ and  $\Delta m^2_{21}$ to their
  best fit values.  }
\end{center}
\end{figure}

\begin{figure}
\begin{center}
\includegraphics[height=8cm,keepaspectratio]{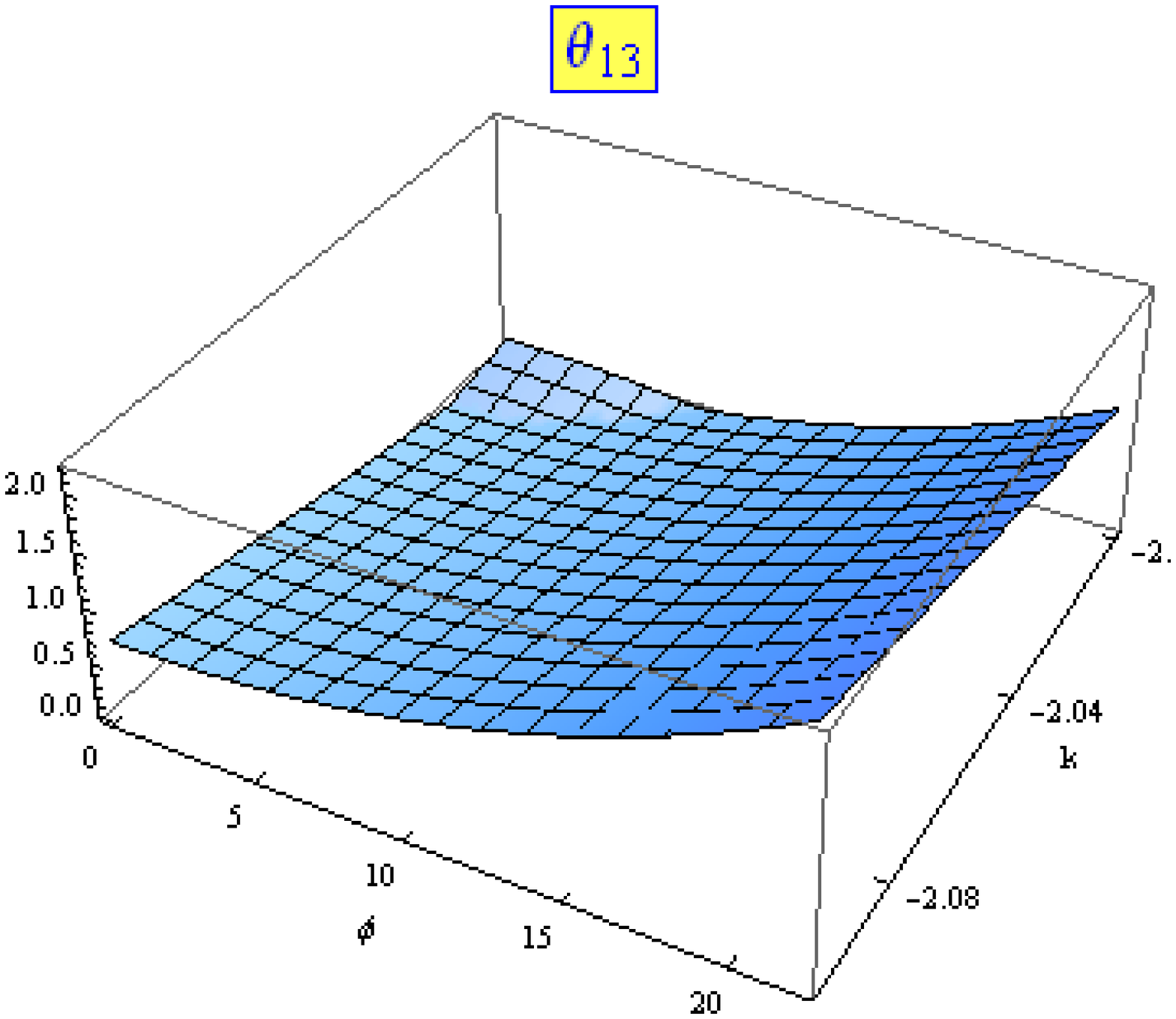}
\caption{\label{th13c13} Plot of   $\theta_{13}$  with respect to
  $k$ and  $\phi$ for the breaking of $A_4$ in ``13'' element of $m_D$. We keep
  $\Delta m^2_{32}$ and  $\Delta m^2_{21}$ to their
  best fit values.  }
\end{center}
\end{figure}
\begin{eqnarray}
\e=\frac{3k\left[R(2\cos\phi-k)(1+k^2+2k\cos\phi)-(2\cos\phi+k)(1+k^2-2k\cos\phi)\right]}{4\left[R(1+k^2-2k\cos\phi)(1+k^2+2k\cos\phi)+(2+k^2+2k\cos\phi)(1+k^2-2k\cos\phi)\right]}.\nonumber\\ 
\label{epc13} 
\end{eqnarray} 
Now using the relation of $\Da$ with the parameters Eq. \ref{msdc13} we get the expression for $m_0^2$:
\begin{eqnarray}
 m_0^2=\frac{9(1+k^2-2k\cos\phi)\Da}{9k(2\cos\phi-k)-12\e(1+k^2-2k\cos\phi)}.
\label{m013c} 
\end{eqnarray} 
where $\e$ is in the form of Eq. \ref{epc13}. Thus,  $\e$ and $m_0^2$ depend on the parameters $k$, $\phi$ and experimentally known $R$. Extraction of $k$ and $\phi$ from other two known mixing angles is little bit difficult. Rather we have plotted $\theta_{12}$ and $\theta_{23}$ with respect to $k$ and $\phi$ and obtain the restriction on the parameter space of $k$ and $\phi$. From the expression of $\sin\theta_{12}$ in Eq. \ref{t12c13} we are seeing that there is a factor $k+2\cos\phi$ in the denominator. For $k>-2$ there will be a $\phi$ for which the quantity $k+2\cos\phi$ becomes zero. Hence, we should keep $k<-2$. Again the factor $k+2\cos\phi$ should be small to ensure that $\e$ is also small. It justifies our whole analysis because we consider first order perturbation as we considered symmetry $A_4$ remains approximate. We consider the range $-2.1<k<-2.0$ and $0<\phi<22^\circ$. From Fig. \ref{th12c13} we see that  $\theta_{12}$ changes  from the tri-bi maximal value $35.6^\circ$ to  $31^\circ$. Near $\phi=10^\circ$ it crosses the best  fit value $34^\circ$. We have plotted $\theta_{23}$ in Fig. \ref{th23c13}. The variation of $\theta_{23}$ is from $47^\circ$ to $44.8^\circ$  for  the same range of $k$ and $\phi$. The best fit value $45^\circ$ is remain within range of variation and it is in the low $k$ low $\phi$ region. The plot of $\theta_{13}$ is in Fig. \ref{th13c13}. Value of $\theta_{13}$ remains within $2^\circ$ for the same range of $k$ and $\phi$ and the model predicts $\theta_{13}$ is very small but non-zero. Question may arise whether such small value of $\theta_{13}$ can generate observable CP violation or not. 

Keeping all those constraints in view next we explore the parameter
space of CP violation parameter $J_{\rm CP}$. The parameter
$J_{\rm CP}$ defined as \cite{branco1}
\begin{eqnarray}
J_{CP} = 
  \frac{1}{8}\sin2\theta_{12}\sin2\theta_{23}\sin2\theta_{13}
  \cos\theta_{13}\sin\delta_{CP}
   = \frac{Im[h'_{12}h'_{23}h'_{31}]}
   {\Delta m^2_{21}\Delta m^2_{31}\Delta m^2_{32}}
\label{jcp1}
\end{eqnarray}
where $h'= M_\nu M_\nu^\dagger$, $\delta_{CP}$ is Dirac phase. This $J_{\rm
  CP}$ is associated with CP violation in neutrino oscillation and is
directly related to Dirac phase of mixing matrix. 
\begin{figure}
\begin{center}
\includegraphics[height=8cm,keepaspectratio]{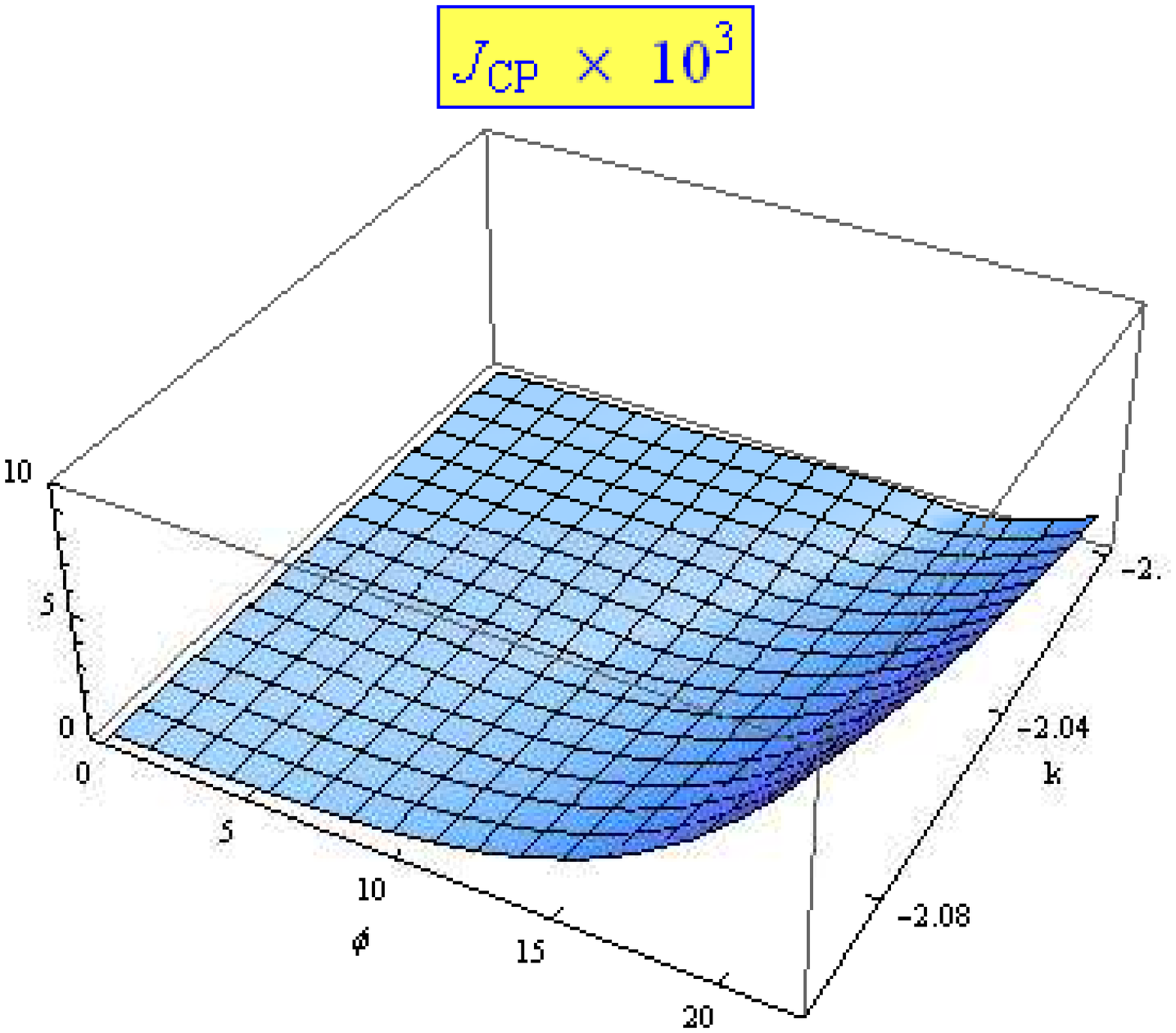}
\caption{\label{fjcp1} Plot of $J_{\rm CP}$ with respect to $k$
  and $\phi$ for the breaking of $A_4$ in 13 element of $m_D$ in the unit of $10^{-3}$. We keep
  $\Delta m^2_{32}$ and  $\Delta m^2_{21}$ to their
  best fit values.  }
\end{center}
\end{figure}
%
\begin{figure}
\begin{center}
\includegraphics[height=8cm,keepaspectratio]{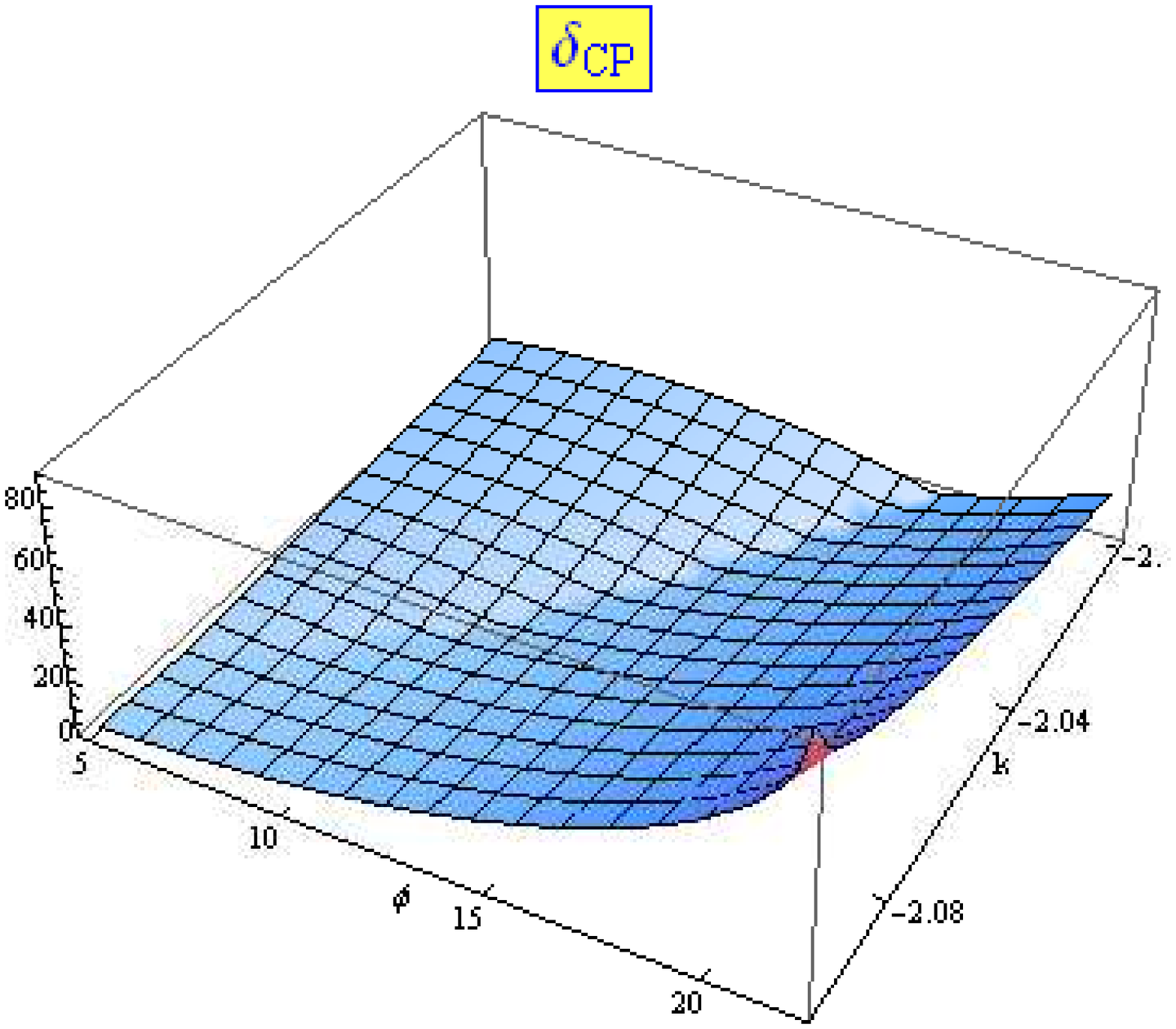}
\caption{\label{dcp13} Plot of $\delta_{\rm CP}$ with respect to $k$
  and $\phi$ for the breaking of $A_4$ in ``13'' element of $m_D$. We keep
  $\Delta m^2_{32}$ and  $\Delta m^2_{21}$ to their
  best fit values.  }
\end{center}
\end{figure}
%
The from Eq.\ (\ref{mnuc}) we can express $M_\nu$ matrix in terms of $\e$ and $m_0$ and $k$ and $\phi$. With the expressions for $\e$ and $m_0$ we can easily obtain the $M_\nu$ and hence $h'=M_\nu M_\nu^\dagger$ completely in terms of $k$ and $\phi$. Hence, similar to the mixing angles $J_{CP}$ will be  also only function of $k$ and  $\phi$. We have plotted  $J_{CP}$ in Fig. \ref{fjcp1} where the values are normalized by a factor $10^{-3}$.
For the same range of $k$ and  $\phi$ the model predicts $J_{CP}$ up to the  order of $10^{-2}$ which is appreciable to observe through the forthcoming experiments. Inverting the expression of $J_{CP}$, the phase $\delta_{CP}$ is extracted in terms of  $k$ and  $\phi$ and it is plotted in Fig. \ref{dcp13}. We see that the value of $\delta_{CP}$ is large upto $90^\circ$ and, thereby, compensates small $\theta_{13}$ effect in $J_{CP}$ and makes it observable size. One important discussion we have to make about the range of $\phi$ and $k$. One can ask why we are  keeping ourselves small range of those parameters where larger $\phi$ can enhance the $\theta_{13}$ and $J_{CP}$. We have studied that the larger value of $\phi$ and also $k$ in negative become responsible for breaking the analytic bound $\sin\delta_{CP}<1$. So, we keep ourselves in shrinked parameter space which keep $\theta_{12}$ and $\theta_{23}$ in exceptionally good values according to the experiment and also can able to generate observable CP violation instead of small $\theta_{13}$. Another thing we want to point out that negative value of $\phi$ equally acceptable as far as it is small. It could not change mixing angles because their expressions depend on $\phi$ through $\cos\phi$ and $\sin^2\phi$. Only $J_{CP}$ and $\delta_{CP}$ will change in sign which are unsettled according to the experiments.  

\begin{figure}
\begin{center}
\includegraphics[height=8cm,keepaspectratio]{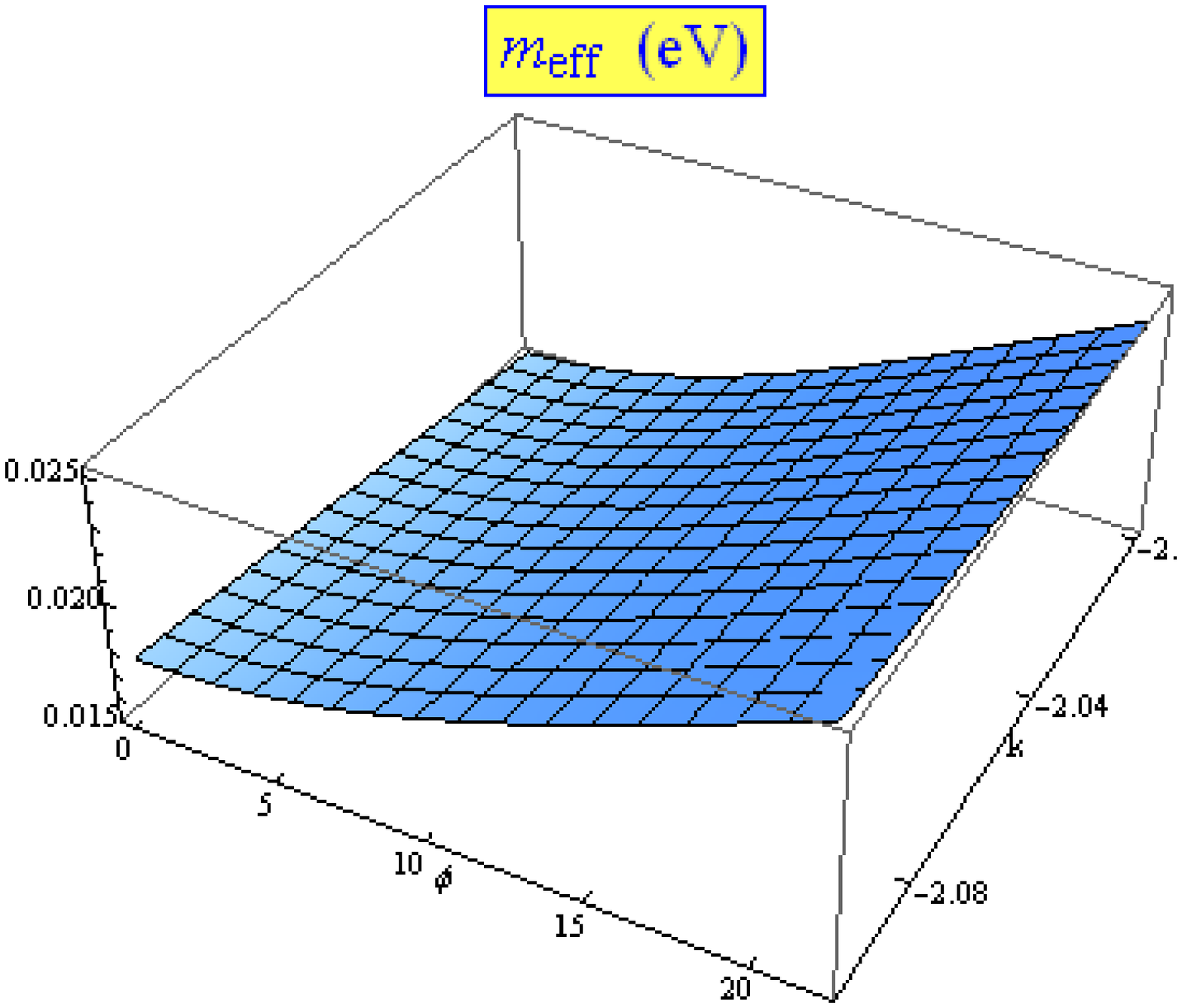}
\caption{\label{mee13} Plot of $m_{\rm eff}=\left|(M_\nu)_{\rm ee}\right|$ with respect to $k$
  and $\phi$ for the breaking of $A_4$ in ``13'' element of $m_D$ . We keep
  $\Delta m^2_{32}$ and  $\Delta m^2_{21}$ to their
  best fit values.  }
\end{center}
\end{figure}

At the end we want to check whether the range of $k$ and  $\phi$ can satisfy the double beta decay bound $m_{\rm eff}=\left|(M_\nu)_{ee}\right|\le 0.89$ eV. In our model expression for this quantity is :
\begin{eqnarray}
m_{\rm eff}=\left|(M_\nu)_{ee}\right|=\frac{m_0}{3}\times\sqrt{\frac{9+k^2(1+2\e)^2+6k(1+2\e)\cos\phi}{1+k^2+2k\cos\phi}}
\label{mnee} 
\end{eqnarray} 
 and it will be also only function of $k$ and $\phi$. We plot this in Fig. \ref{mee13} and it remains well below the experimental upper bound. 

Again we want to discuss about the mass pattern. Throughout our whole analysis in real as well as complex case we keep $\Ds$ and $\Da$ to their best fit value and take the negative sign of $\Da$. It corresponds to so called inverted ordering of neutrino mass. It is the feature near $k=-2$. It is necessary to keep $m_0^2>0$. Why we so fond of this region of $k$ instead of region $k=1$ which can give the normal hierarchical mass spectrum. The reason is that the inverted ordering corresponds to the light neutrino mass scale $m_0\simeq \Da$ where $m_0\simeq \Ds$ for normally ordered mass spectrum. So, from the point of view of observable CP violation, it is inevitable to choose larger value of $m_0$ because $J_{CP}\propto m_0^6$. So, inverted hierarchical mass spectrum compatible with the observable CP violation. Now we extend our study of this model to leptogenesis. We want to see whether we can have appreciable leptogenesis compatible with baryon asymmetry for the same parameter space $k$ and $\phi$ after successful low energy data analysis for the feasible value of scale $M_0$. 

%
\subsection{Leptogenesis}
After successful predictions of low energy neutrino data we want to see whether this model can generate non-zero lepton-asymmetry  with proper size and sign to describe baryon asymmetry. We keep the same right handed neutrino mass matrix  $M_R$ as before. The change only appear in Dirac type Yukawa coupling and hence in $m_D$ also. The diagonalisation of $M_R$ gives 
\begin{eqnarray}
M_R^D=U_{TB}^\dagger M_R U_{TB}^*={\rm diag}(|A+De^{i\phi}|e^{i2\alpha},~|A|e^{i2\lambda},~|De^{i\phi}-A|e^{i2\gamma})
\label{mrd} 
\end{eqnarray} 
and hence the masses of the right handed neutrino are
\begin{eqnarray}
 M_1=|A+De^{i\phi}|&=&M_0\sqrt{1+k^2+2k\cos\phi}\qquad
M_2=|A|=M_0\nonumber\\
M_3&=&|De^{i\phi}-A|=M_0\sqrt{1+k^2-2k\cos\phi}
\label{mrev} 
\end{eqnarray} 
and the phases are
\begin{eqnarray}
 \tan{2\alpha}=\frac{k\sin\phi}{1+k\cos\phi}\qquad \tan{2\lambda}=0\qquad\tan{2\gamma}=\frac{k\sin\phi}{k\cos\phi-1}.
\label{majp} 
\end{eqnarray} 
The explicit form of diagonalising matrix is
\begin{eqnarray}
V=U_{TB}U_P=\pmatrix{\sqrt{\frac{2}{3}}& \sqrt{\frac{1}{3}} &0\cr
                    -\sqrt{\frac{1}{6}}& \sqrt{\frac{1}{3}} & -\sqrt{\frac{1}{2}}\cr
                    -\sqrt{\frac{1}{6}}&\sqrt{\frac{1}{3}}& \sqrt{\frac{1}{2}}}\pmatrix{e^{i\alpha}&0&0\cr
                    0&e^{i\lambda}&0\cr
                    0&0&e^{i\gamma}}
\label{mutb} 
\end{eqnarray} 
where the expressions for the phases are given in Eq.\ (\ref{majp}). The $m_D$ matrix is no longer diagonal after explicit breaking of $A_4$ symmetry. In the mass basis of right handed neutrino the modified Dirac mass term is
$m_D'=m_DV^*$. Hence the relevant matrix  for describing Leptogenesis is :					
\begin{eqnarray}
 h=m_D'^{\dagger}m_D'=\frac{f^2v_u^2}{2}\pmatrix{1-2\e/3& \e e^{ i(\alpha-\lambda)}/(3\sqrt{2})&\e e^ {i(\alpha-\gamma)}/\sqrt{3}\cr
                \e e^ {-i(\alpha-\lambda)}/(3\sqrt{2})&1+2\e/3&\e e^{i(\lambda-\gamma)}/\sqrt{6}\cr
                 \e e^ {-i(\alpha-\gamma)}/\sqrt{3}&\e e^{-i(\lambda-\gamma)}/\sqrt{6}&1}.
\label{hc13} 
\end{eqnarray} 
To calculate lepton asymmetry as in  Eq.\ (\ref{vertex}) and Eq.\ (\ref{self}) we need to calculate following quantities from matrix $h$:
\begin{eqnarray}
&&{\rm Im}(h_{21}^2)=- {\rm Im}(h_{12}^2)=-\frac{f^4\e^2v_u^4\sin{2(\alpha-\lambda)}}{72}=-\frac{f^4\e^2v_u^4}{72}\times\frac{k\sin\phi}{\sqrt{1+k^2+2k\cos\phi}}\nonumber\\
&&{\rm Im}(h_{32}^2)=- {\rm Im}(h_{23}^2)=-\frac{f^4\e^2v_u^4\sin{2(\lambda-\gamma)}}{12}=\frac{f^4\e^2v_u^4}{24}\times\frac{k\sin\phi}{\sqrt{1+k^2-2k\cos\phi}}\nonumber\\
&&{\rm Im}(h_{31}^2)=- {\rm Im}(h_{13}^2)=-\frac{f^4\e^2v_u^4\sin{2(\alpha-\gamma)}}{12}=\frac{2f^4\e^2v_u^4}{12}\times\frac{k\sin\phi}{\sqrt{(1+k^2)^2-4k^2\cos^2\phi}}.\nonumber\\
\label{imh} 
\end{eqnarray} 
Calculating $x_{ij}=M_j^2/M_i^2$ from Eq. \ref{mrev}, $h_{ii}$ from Eq.\ (\ref{hc13}) and taking ${\rm Im}(h_{ij}^2)$ from Eq.\ (\ref{imh}) we calculate the self energy part of lepton asymmetry from Eq.\ (\ref{self}) and vertex part of lepton asymmetry from Eq.\ (\ref{vertex}). Adding both we obtain the following decay asymmetry of right handed neutrinos for all three generations
\begin{eqnarray}
&&\varepsilon_1=\frac{M_0m_0\e^2k\sin\phi}{72\pi v_u^2(1+k^2+2k\cos\phi)}\times\left[
\frac{1+2k^2+4k\cos\phi}{k(k+2\cos\phi)}-\frac{3(1+k^2+6k\cos\phi)}{k\cos\phi}\right .\nonumber\\
&&-\left .\frac{(2+k^2+2k\cos\phi)\ln(2+k^2+2k\cos\phi)}{1+k^2+2k\cos\phi}+\frac{24(1+k^2)}{1+k^2+2k\cos\phi}\times\ln\frac{2(1+k^2)}{1+k^2-2k\cos\phi}\right].\nonumber\\
\label{las1} 
\end{eqnarray} 
\begin{eqnarray}
\varepsilon_2&=&-\frac{M_0m_0\e^2k\sin\phi}{72\pi v_u^2}\times\left[
\frac{k^2+2k\cos\phi-1}{k(k+2\cos\phi)}+\frac{3(1-k^2+2k\cos\phi)}{k(2\cos\phi-k)}-(2+k^2+2k\cos\phi)\right .\nonumber\\
&&\left .\times\ln\left\{\frac{2+k^2+2k\cos\phi}{1+k^2+2k\cos\phi}\right\}-3(2+k^2-2k\cos\phi)\times\ln\left\{\frac{2+k^2-2k\cos\phi}{1+k^2-2k\cos\phi}\right\}\right] 
\label{las2} 
\end{eqnarray} 
\begin{eqnarray}
&&\varepsilon_3=\frac{M_0m_0\e^2k\sin\phi}{24\pi v_u^2(1+k^2-2k\cos\phi)}\times\left[
\frac{6k\cos\phi-k^2-1}{k\cos\phi}+\frac{1+2k^2-4k\cos\phi}{k(k-2\cos\phi)}\right .\nonumber\\
&&-\left .\frac{2+k^2-2k\cos\phi}{1+k^2-2k\cos\phi}\times\ln(2+k^2-2k\cos\phi)-\frac{8(1+k^2)}{1+k^2-2k\cos\phi}\times\ln\frac{2(1+k^2)}{1+k^2+2k\cos\phi}\right]\nonumber\\
\label{las3} 
\end{eqnarray} 
CP asymmetry parameters $\varepsilon_i$ are related
to the leptonic asymmetry parameters through $Y_L$ as
\cite{Pilaftsis1,Nielsen:2001fy,Pilaftsis:2003gt}
\begin{eqnarray}
Y_L\equiv\frac{n_L-{\bar n}_L}{s}=\sum_i^3\frac{\varepsilon_i\kappa_i}{g_{*i}}
\label{leptasym}
\end{eqnarray}
where $n_L$ is the lepton number density, ${\bar n}_L$ is the
anti-lepton number density, $s$ is the entropy density, $\kappa_i$ is
the dilution factor for the CP asymmetry $\varepsilon_i$ and $g_{*i}$
is the effective number of degrees of freedom \cite{Roos:1994fz} at
temperature $T=M_i$. Value of $g_{*i}$ in the SM with three right
handed Majorana neutrinos and one extra Higgs doublet is $116$. The baryon asymmetry $Y_B$
produced through the sphaleron transmutation of $Y_L$ , while the
quantum number $B-L$ remains conserved, is given by
\cite{Harvey:1990qw}
\begin{eqnarray}
Y_B=\frac{\omega}{\omega-1}Y_L \qquad {\rm with} \qquad \omega =\frac{8N_F+4N_H}{22N_F+13N_H},
\label{barasym}
\end{eqnarray}
where $N_F$ is the number of fermion families and $N_H$ is the number
of Higgs doublets. The quantity $\omega=8/23$ in Eq.\ (\ref{barasym})
for SM with two Higgs doublet. Now we introduce the relation between $Y_B$ and $\eta_B$,
where $\eta_B$ is the baryon number density over photon number density
$n_\gamma$. The relation is \cite{Barger}
\begin{eqnarray}
\eta_B=\left.\frac{s}{n_\gamma}\right|_0Y_B=7.0394Y_B,
\label{yetar}
\end{eqnarray}
where the zero indicates present time. Now using the relations in Eqs.\
(\ref{leptasym},\ref{barasym}, \ref{yetar}), $\omega=8/23$ and $g_{*i}=116$ we have
\begin{eqnarray}
\eta_B=-3.23\times 10^{-2}\sum_i\varepsilon_i\kappa_i.
\label{etab}
\end{eqnarray}
This dilution factor $\kappa_i$ approximately given by \cite{Nielsen:2001fy,kolb,Giudice:2003jh}
\begin{eqnarray}
\kappa_i\simeq\frac{0.3}{K_i(\ln K_i)^{3/5}} \qquad {\rm with}\qquad  K_i=\frac{\Gamma_i}{H_i},
\label{kppa}
\end{eqnarray}
where $\Gamma_i$ is the decay width of $N_i$ and $H_i$ is Hubble
constant at $T=M_i$. Their expressions are
\begin{eqnarray}
\Gamma_i=\frac{h_{ii}M_i}{4\pi v_u^2}\qquad {\rm and} \qquad H_i=1.66\sqrt{g_{*i}}\frac{M_i^2}{M_P},
\label{gh}
\end{eqnarray}
where $v_u=v\sin\beta$, $v=246$GeV and $M_P=1.22\times 10^{19}$GeV. Thus we have 
\begin{eqnarray}
K_i=\frac{M_Ph_{ii}}{1.66\times 4\pi\sqrt{g_{*i}}v_u^2M_i}.
\label{K}
\end{eqnarray}
%
For our model $K_1$, $K_2$ and $K_3$ are
\begin{eqnarray}
&&K_1= \frac{M_Ph_{11}}{1.66\times 4\pi\sqrt{g_{*1}}v_u^2M_1}=N\times\frac{m_0}{v_u^2}\times\frac{(1-\frac{2\e}{3})}{(1+k^2+2k\cos\phi)^{1/2}}\nonumber\\
&&K_2=\frac{M_Ph_{22}}{1.66\times 4\pi\sqrt{g_{*2}}v_u^2M_2}=N\times \frac{m_0}{v_u^2}\times (1+\frac{2\e}{3})\nonumber\\
&&K_3=\frac{M_Ph_{33}}{1.66\times 4\pi\sqrt{g_{*3}}v_u^2M_3}=N\times\frac{m_0}{v_u^2}\times\frac{1}{(1+k^2-2k\cos\phi)^{1/2}}\nonumber\\
\label{k1k2k3} 
\end{eqnarray} 
\begin{figure}
\begin{center}
\includegraphics[height=8cm,keepaspectratio]{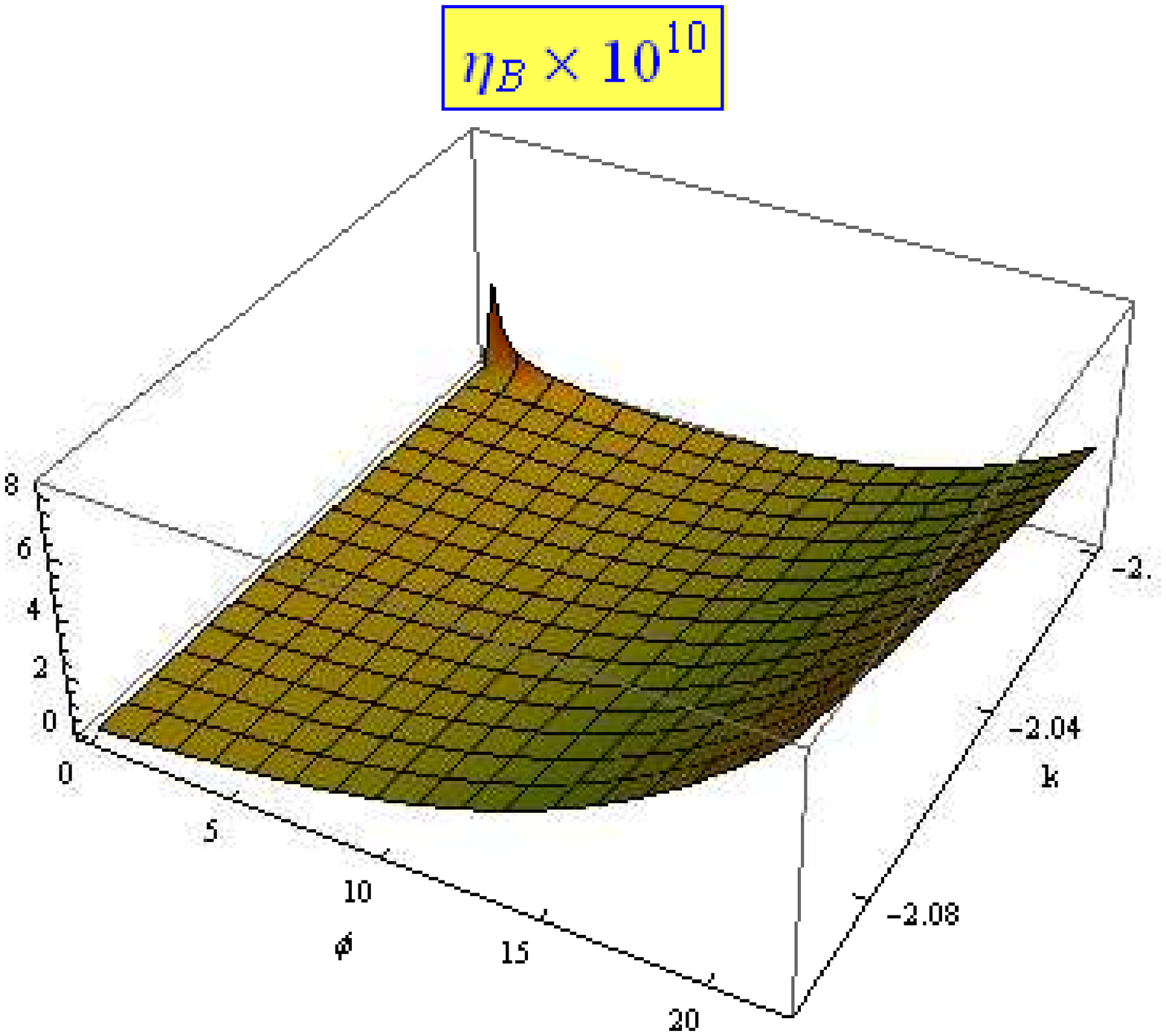} 
\caption{\label{basymp1} Plot of baryon asymmetry $\eta_{\rm B}$ in unit of $10^{-10}$ with respect to $k$
  and $\phi$ for the breaking of $A_4$ in ``13'' element of $m_D$ . We keep
  $\Delta m^2_{32}$ and  $\Delta m^2_{21}$ to their
  best fit values and have plotted for mass scale of right handed neutrino $M_0=2\times 10^{13}$ GeV}
\end{center}
\end{figure}
\begin{figure}
\begin{center}
\includegraphics[height=8cm,keepaspectratio]{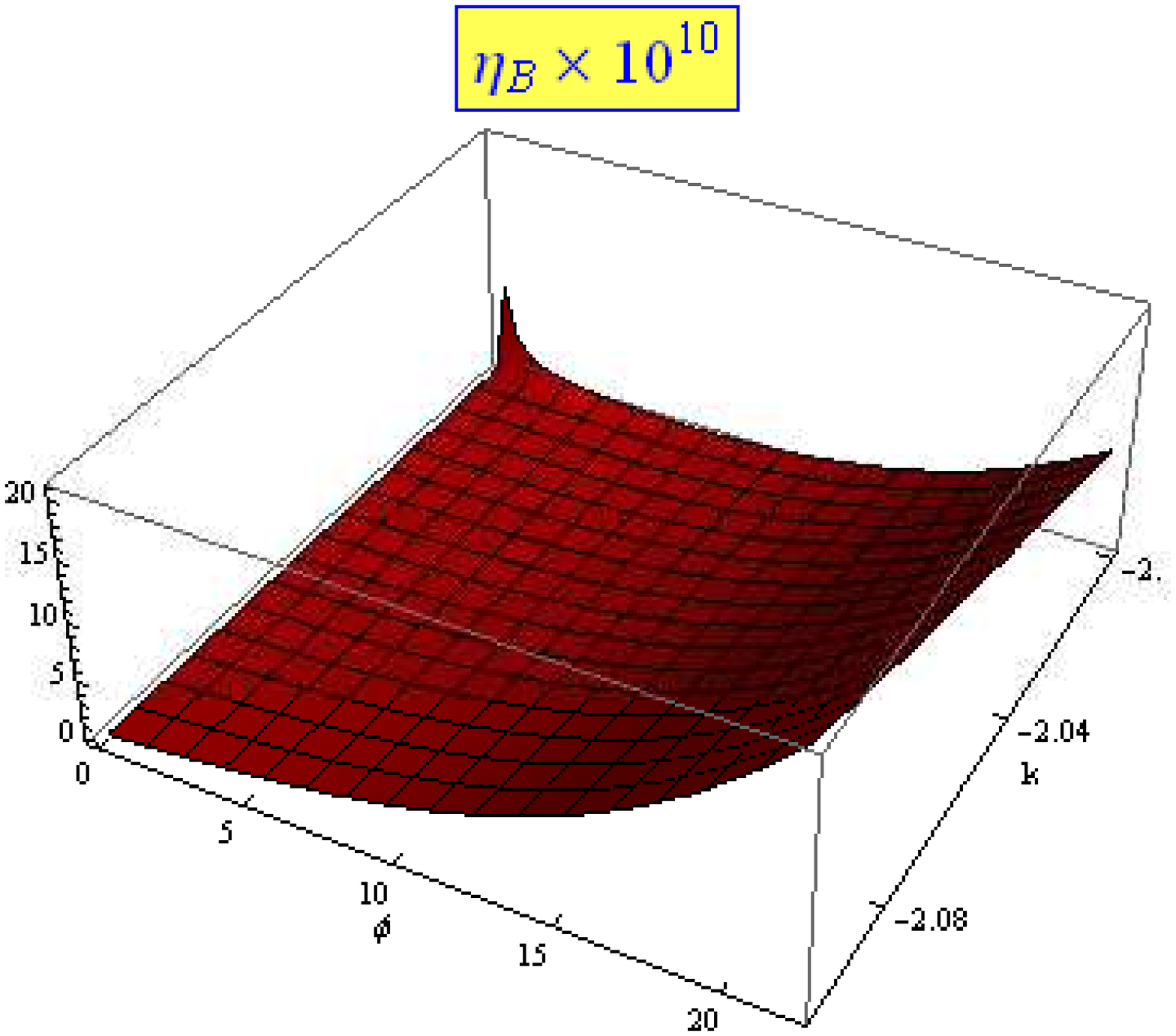} 
\caption{\label{basymp2} Plot of baryon asymmetry $\eta_{\rm B}$ in unit of $10^{-10}$ with respect to $k$
  and $\phi$ for the breaking of $A_4$ in ``13'' element of $m_D$ . We keep
  $\Delta m^2_{32}$ and  $\Delta m^2_{21}$ to their
  best fit values and have plotted for mass scale of right handed neutrino $M_0=5\times 10^{13}$ GeV }
\end{center}
\end{figure}
\begin{figure}
\begin{center}
\includegraphics[height=8cm,keepaspectratio]{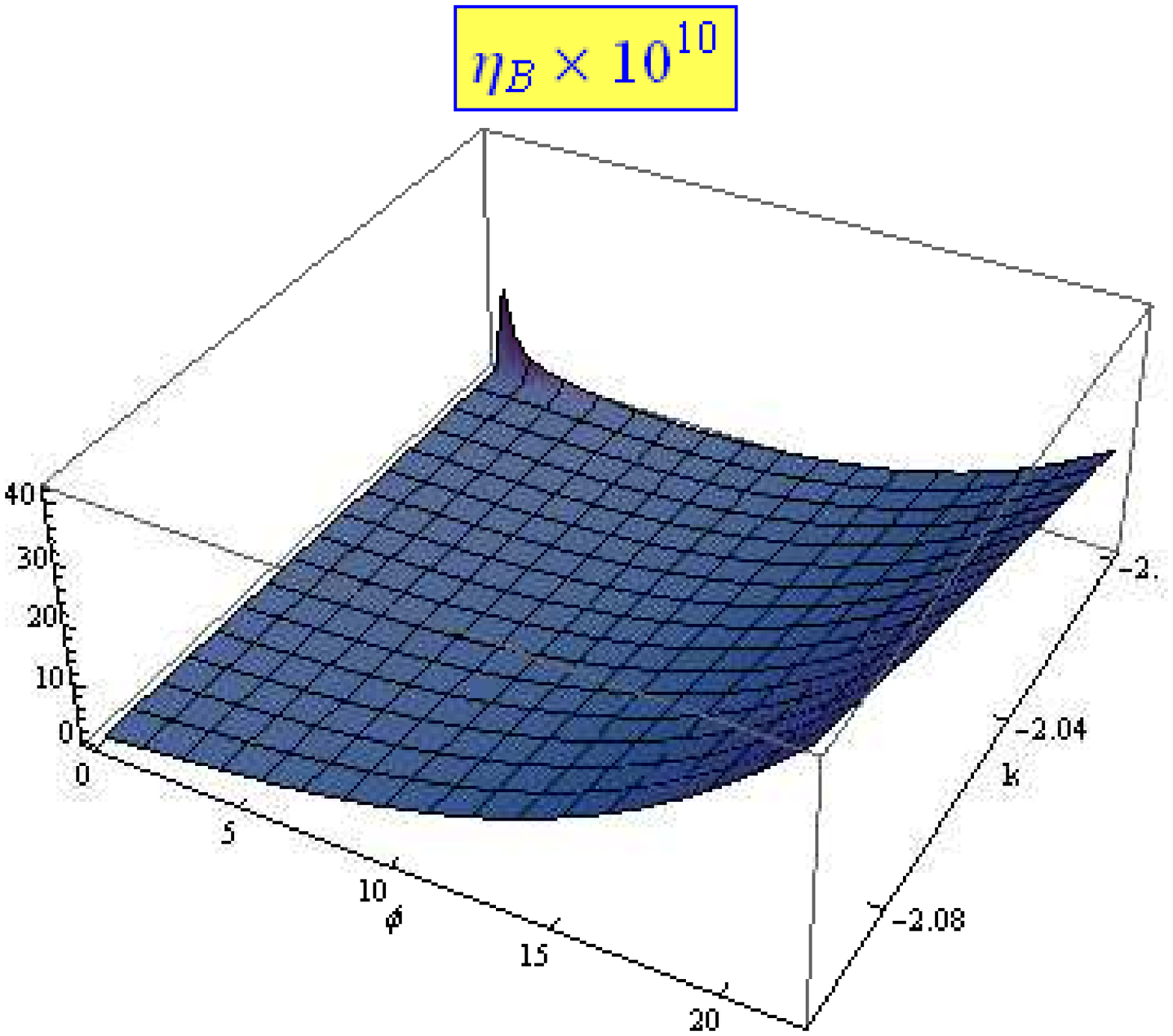} 
\caption{\label{basymp3} Plot of baryon asymmetry $\eta_{\rm B}$ in unit of $10^{-10}$ with respect to $k$
  and $\phi$ for the breaking of $A_4$ in ``13'' element of $m_D$ . We keep
  $\Delta m^2_{32}$ and  $\Delta m^2_{21}$ to their
  best fit values and have plotted for mass scale of right handed neutrino  $M_0=10^{14}$ GeV }
\end{center}
\end{figure}
\begin{figure}
\begin{center}
\includegraphics[height=8cm,keepaspectratio]{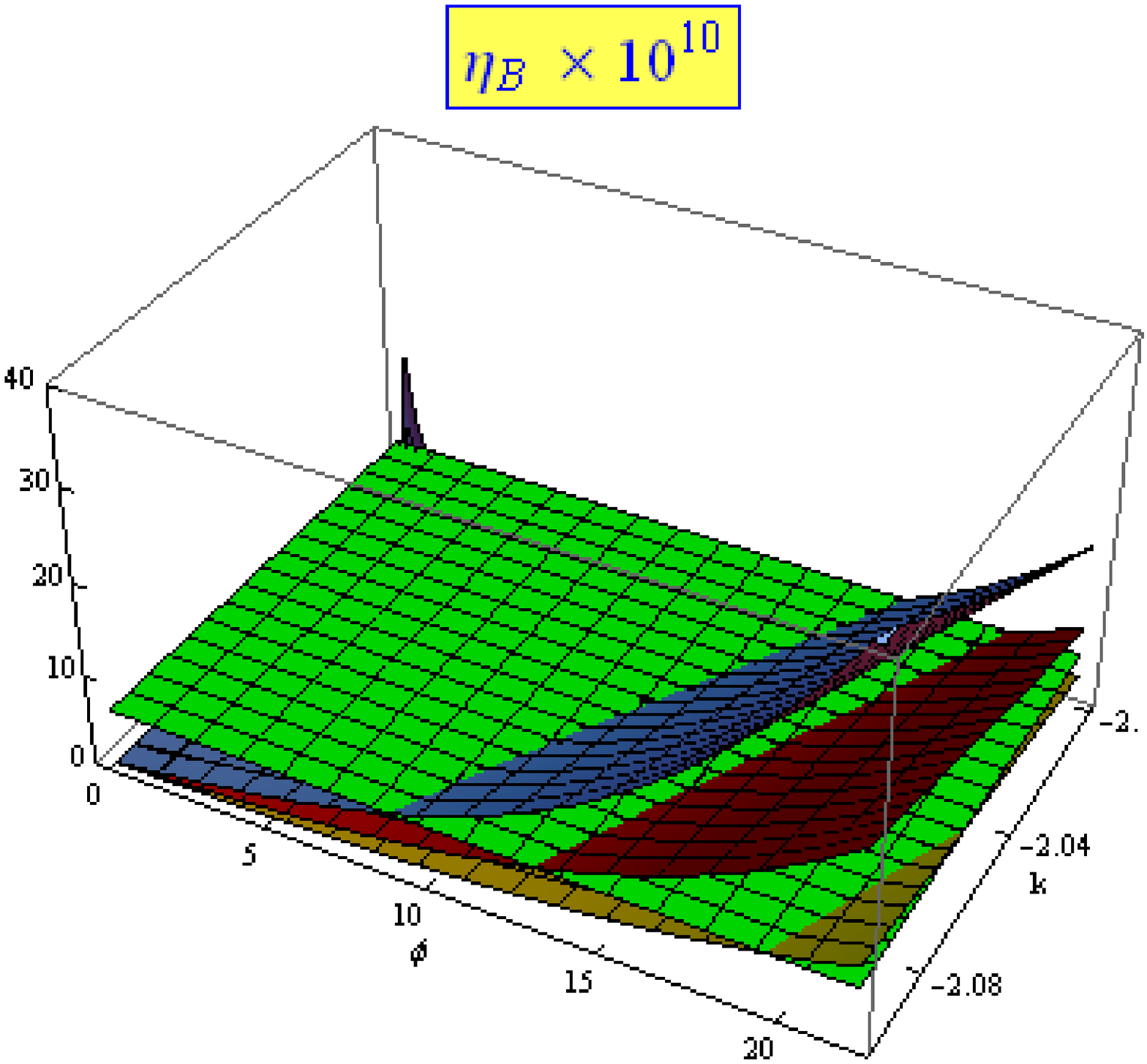}
\caption{\label{basym0} We combine all three plots of baryon asymmetry for three right handed neutrino mass scale along with the WMAP value of baryon asymmetry $6\times 10^{-10}$ which is the plane surface.}
\end{center}
\end{figure}
%
where $N=\frac{M_P}{1.66\times 4\pi\sqrt{g_{*}}}$, $g_{*}= g_{*1}\approx g_{*2}\approx g_{*3}\approx 116$. Using Eq.\ (\ref{k1k2k3}) into Eq.\ (\ref{kppa}) and from the expression of $\varepsilon_i$ we can say that apart from logarithmic factor, $\kappa_i\propto v_u^2/m_0$ and $\varepsilon_i\propto m_0/v_u^2$. So, baryon asymmetry will be independent of $m_0$ and $v_u$. They only appear through logarithmic factor in $\kappa_i$. We consider $\tan\beta\approx 2.5$.  Substituting $m_0$ from Eq.\ (\ref{m013c}), $\e$ from Eq.\ (\ref{epc13}) and considering $M_0$ some specific value into the expressions of $\varepsilon_i$, $\kappa_i$ we can have baryon asymmetry as function of $k$ and $\phi$ only. In Fig. \ref{basymp1}, Fig. \ref{basymp2} and Fig. \ref{basymp3} we have plotted $\eta_B$ as function of $k$ and $\phi$ in the unit of $10^{-10}$ for three $M_0$ values, $M_0=2\times 10^{13}$ GeV,  $M_0=5\times 10^{13}$ GeV and $M_0=10^{14}$ GeV rspectively. We have seen that the experimental value of $\eta_B\simeq 6.0\times 10^{-10}$ is obtainable in our model within the same range of $\kappa$ and $\phi$ as in the low energy case. To see more explicitly the baryon asymmetry plots we combine all three plots along with the observed baryon asymmetry value $6\times 10^{-10}$ which corresponds the plane  surface in Fig. \ref{basym0}. The observed WMAP value of the baryon asymmetry curve intersect the lower curve (for $M_0=2\times 10^{13}$ GeV) near the boundary of $\kappa$ and $\phi$ variation. So, lower value of $M_0$ could not generate observable baryon asymmetry. In the intersection region $\theta_{12}\simeq 32^\circ$ and $\theta_{23}\simeq 47^\circ$. If we allow that much of variation of $\theta_{12}$ and $\theta_{23}$, we can have large low energy CP violation as well as baryon asymmetry with proper size and sign with $M_0=2\times 10^{13}$ GeV which is near the upper bound of right handed neutrino mass scale for generation of lepton as well as baryon asymmetry. If we more relax, we can easily see that the intersection of experimental and theoretical curve for $M_0=5\times 10^{13}$ GeV and $M_0=10^{14}$ GeV is in the lower value of $k$ and $\phi$ where well known neutrino mixing angles are more closer to their best fit value. 

Let us give a close look to the plot of $\eta_B$ in Fig. \ref{basymp1}, Fig. \ref{basymp2} and Fig. \ref{basymp3} near very low $\phi$ and $k\simeq -2$ region. This is the region where $\eta_B $ become very large. From the expression of $\varepsilon_1$ and $\varepsilon_2$ it is clear that $\varepsilon_1$ and $\varepsilon_2$  both are singular at $k+2\cos\phi= 0$. This corresponds to equality of the masses $M_1=M_2$ or $x_{12}=x_{21}=1$. This singularity can be avoided  considering finite decay width of right  handed neutrinos. We can able to maximize $\varepsilon_1$ and $\varepsilon_2$ and hence $\eta_B$ using resonant condition $M_1-M_2\simeq \Gamma_{1,2}/2$. But, as we have already obtained the observed baryon asymmetry without resonance, it is not necessary to think about so finely tuned condition. Again in the region where the resonant condition is applicable, the $J_{CP}$ is miserably small to observe through any experiments.

\subsection{Final Analysis}
\label{sec:fan}
\begin{figure}
\begin{center}
 \includegraphics[height=6cm,keepaspectratio]{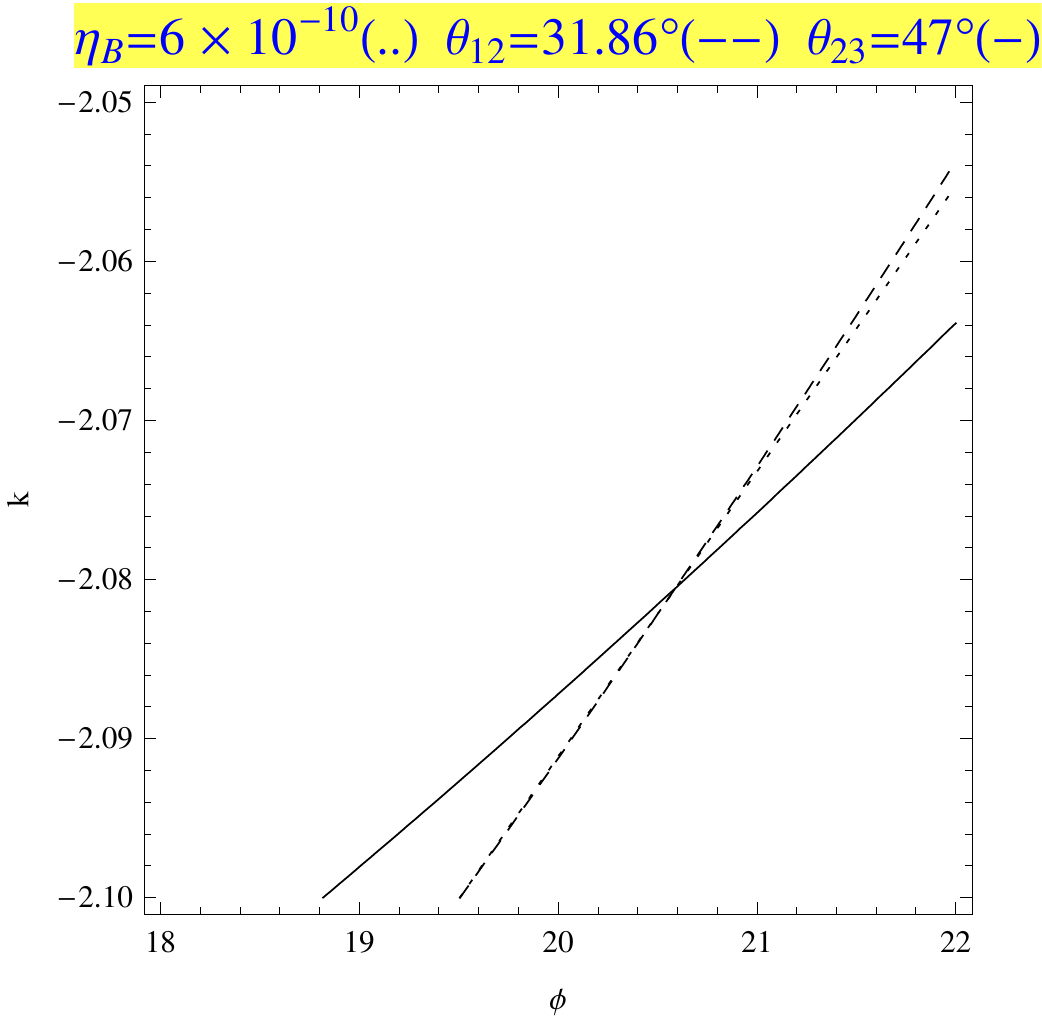}
\caption{\label{fig:contplot1} Contour plot of baryon asymmetry $\eta_{\rm B}$, $\theta_{12}$ and, $\theta_{23}$ in $\phi$-$k$ plane for  $\Delta m^2_{32}$ and  $\Delta m^2_{21}$ to their
  best fit values and for mass scale of right handed neutrino $M_0=2\times 10^{13}$ GeV.}
\end{center}
\end{figure}
\begin{figure}
\begin{center}
 \includegraphics[height=6cm,keepaspectratio]{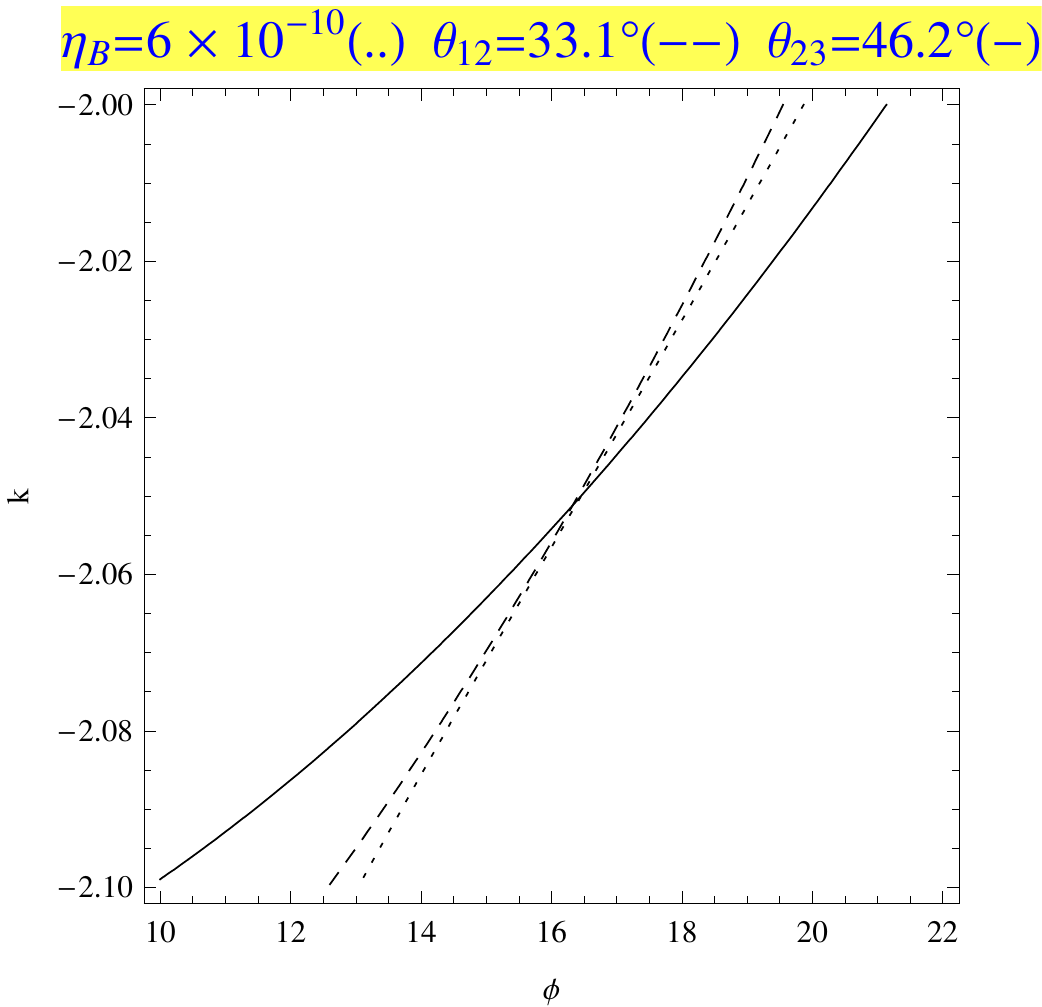}
\caption{\label{fig:contplot2} Contour plot of baryon asymmetry $\eta_{\rm B}$, $\theta_{12}$ and, $\theta_{23}$ in $\phi$-$k$ plane for  $\Delta m^2_{32}$ and  $\Delta m^2_{21}$ to their
  best fit values and for mass scale of right handed neutrino  $M_0=5\times 10^{13}$ GeV.}
\end{center}
\end{figure}
\begin{figure}
\begin{center}
\includegraphics[height=6cm,keepaspectratio]{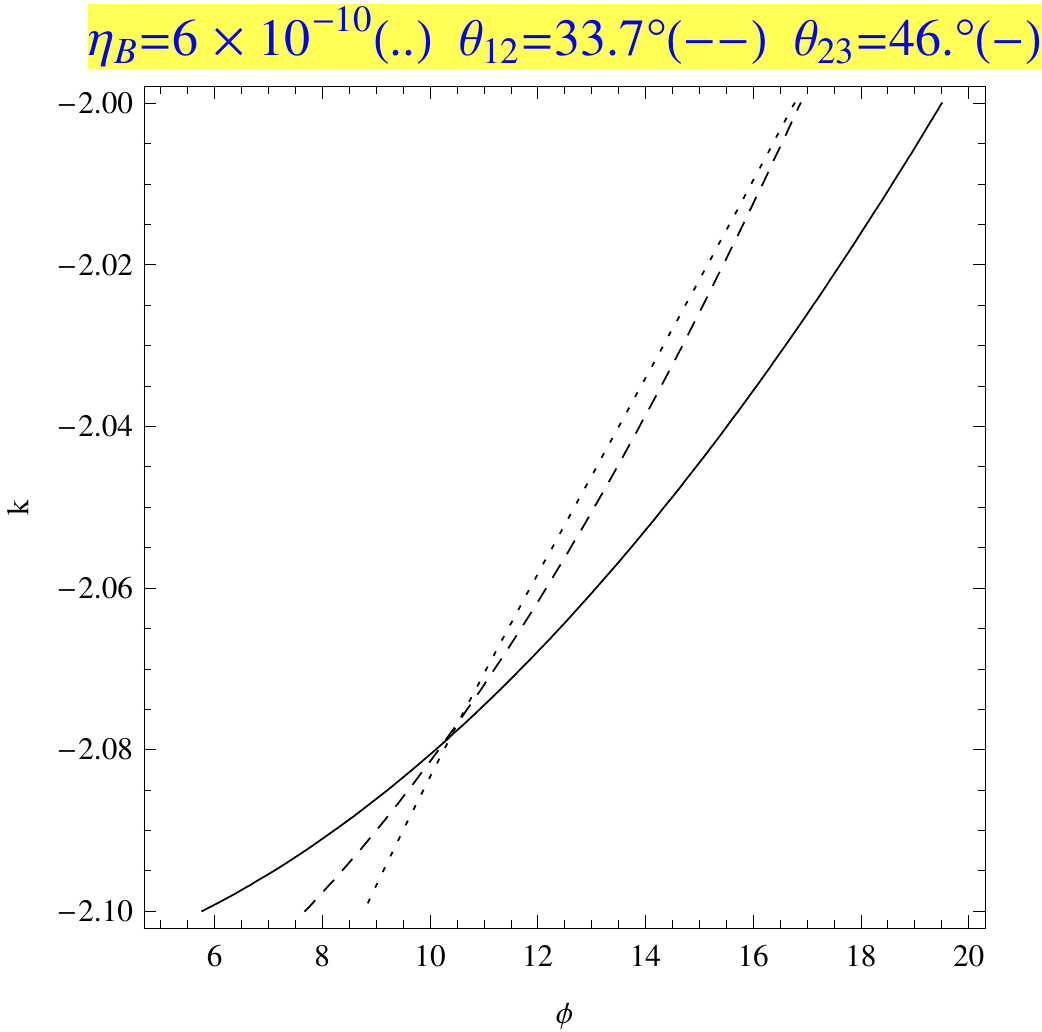}
\caption{\label{fig:contplot3} Contour plot of baryon asymmetry $\eta_{\rm B}$, $\theta_{12}$ and, $\theta_{23}$ in $\phi$-$k$ plane for  $\Delta m^2_{32}$ and  $\Delta m^2_{21}$ to their
  best fit values and for mass scale of right handed neutrino  $M_0=10^{14}$ GeV.}
\end{center}
\end{figure}
We end our analysis with the help of three contour plots of baryon asymmetry $\eta_B=6\times 10^{-10}$ for three scales $M_0=2\times 10^{13}$ GeV, $M_0=5\times 10^{13}$ GeV and, $M_0=10^{14}$ GeV in $\phi$-$k$ plane. We insert the contours of $\theta_{23}$ and $\theta_{12}$ and manage to find the intersection of three contours for some reasonable value of $\theta_{23}$ and $\theta_{12}$.
 
Case(I): $M_0=2\times 10^{13}$ GeV, from Fig. \ref{fig:contplot1}, we are seeing that three contours $\theta_{12}=31.86^\circ$, $\theta_{23}=47^\circ$ and,
$\eta_B=6\times 10^{-10}$ are intersecting at a point ( $20.5$, $-2.082$) in $\phi$-$k$ plane. So, the mixing angles are within nearly $2^\circ$ variation about the best fit values. Obtained $k$, $\phi$ value gives $\theta_{13}=1.78^\circ$, $J_{CP}=0.5\times 10^{-2}$, $\delta_{CP}=46^\circ$, $m_{\rm eff}=0.02423$ eV, $m_0=0.05365$ eV, $\e=-0.121$ and $k+2\cos\phi=-0.208$.

Case (II): $M_0=5\times 10^{13}$ GeV, now from Fig. \ref{fig:contplot2}, we have the intersection of the contour $\eta_B=6\times 10^{-10}$ with the contours $\theta_{12}=33.1^\circ$, $\theta_{23}=46.2^\circ$  at ($16.0$, $-2.07$) in $\phi$-$k$ plane. So, $\theta_{12}$ and $\theta_{23}$ are more closer to their best fit values (nearly $1^\circ$ deviation from their best fit value). At this point we have $\theta_{13}=1^\circ$, $J_{CP}=0.8\times 10^{-3}$, $\delta_{CP}=12^\circ$, $m_{\rm eff}=0.02175$ eV, $m_0=0.0516$ eV, $\e=-0.076$ and $k+2\cos\phi=-0.131$.

Case (III): $M_0=10^{14}$ GeV, from the  Fig. \ref{fig:contplot3}, the contours $\eta_B=6\times 10^{-10}$, $\theta_{12}=33.7^\circ$ and, $\theta_{23}=46.0^\circ$ have crossed in $\phi$-$k$ plane at ($10.5$, $-2.08$). So, higher scale of right handed neutrino mass helps to have very good value of  $\theta_{12}$ and $\theta_{23}$. At the intersection point, we get $\theta_{13}=0.8^\circ$, $J_{CP}=0.23\times 10^{-3}$, $\delta_{CP}=4.5^\circ$, $m_{\rm eff}=0.0194$ eV, $m_0=0.051$ eV, $\e=-0.065$ and $k+2\cos\phi=-0.11$.

The small value of $\e$ compatible with all experimental results. So, we can demand that $A_4$ is an approximate symmetry. 
\subsection{Corelation of CP violating phases in this model}
\label{sec:pr}
In this model everything is determinable in terms of parameter $k$ and $\phi$. Well measured quantities fix the value of those parameters. So, value of the rest of the physical quantities (some of them not so well measured in experiment like $\theta_{13}$, $m_{\rm eff}$ and some of them yet to measure in experiment like $J_{CP}$, $\delta_{CP}$ and the Majorana phases also) are obtainable in this model. Question may arise whether we can have any relations among the phases in this model or not. First of all, there are two kinds phases, low energy and high energy phases. High energy phases are responsible to generate the lepton asymmetry. The low energy phases are responsible for determining low energy leptonic CP violation. Low energy phases are in two type, one is the lepton no preserving CP violating phase $\delta_{CP}$ and another two are the lepton no breaking CP violating phases. In general, all phases are independent, meaning that there are no correlation among the phases between high and low energy sector, and also phases inside a particular sector are not correlated. For three generations of neutrinos, there are three key phases which are responsible for leptogenesis. Those are phases in $h_{12}^2$, $h_{13}^2$, and $h_{23}^2$. From matrix $h$ given in Eq.\ (\ref{hc13}), we have
\begin{eqnarray}
\Phi_1=\arg(h_{12}^2)=2\alpha\qquad \Phi_2=\arg(h_{13}^2)=2(\alpha -\gamma) \qquad \Phi_3=\arg(h_{23}^2)=-2\gamma.
\label{phh}  
\end{eqnarray}  
It leads to the relation,
\begin{eqnarray}
\Phi_2=\Phi_1 + \Phi_3.
\label{hprl}  
\end{eqnarray} 
For a given $k$ and $\phi$ the $\alpha$ and $\gamma$ are known from Eq.\ (\ref{majp}). Hence $\Phi_1$, $\Phi_2$ and $\Phi_3$ are individually determinable phases. But, in this model values of those high energy phases  follow the relation given in Eq.\ (\ref{hprl}).

Now, let us give a fresh look to the leptonic mixing matrix. In Eq.\ (\ref{mnuc}) we have given the tri-bimaximal rotated form of the neutrino mass matrix. Keeping terms upto first order in $\e$ we obtain the diagonalising matrix in the following form,
\begin{eqnarray}
V'=U_{TB}\left(\begin{array}{ccc}
1 & -\e X e^{-i\theta_X} & -\e Y e^{-i\theta_Y} \\
\e X e^{i\theta_X} & 1 & -\e Z e^{-i\theta_Z} \\
\e Y e^{i\theta_Y}  & \e X e^{i\theta_Z} & 1 
\end{array}\right) 
 \label{utbrmm} 
\end{eqnarray} 
where the $X$, $Y$, $Z$ and the associated phases are completely known function of $k$ and $\phi$. An additional phase matrix $V_P$ is needed to make masses of light neutrino real, from Eq.\ (\ref{msevc}) we have the phase matrix
\begin{eqnarray}
V_P=\left(\begin{array}{ccc}
e^{i(\pi/2-\alpha)} & 0 & 0 \\
0 & e^{i\pi/2} & 0 \\
0  & 0 & e^{i(\pi/2-\gamma)} 
\end{array}\right). 
 \label{upm} 
\end{eqnarray} 
To obtain the CKM form of mixing matrix we need to rotate $V'$ by two diagonal phase matrix, let $U_{P1}={\rm diag}(e^{i\alpha_1},~e^{i\alpha_2},~e^{i\alpha_3})$ and  $U_{P2}={\rm diag}(e^{i\beta_1},~e^{i\beta_2},~e^{i\beta_3})$. So, we have
\begin{eqnarray}
V'=U_{P1}^\dagger\left\{U_{P1}V'U_{P2}^\dagger\right\}U_{P2}=U_{P1}^\dagger V_{CKM}U_{P2}
 \label{prot} 
\end{eqnarray} 
Now the with $\theta_{13}$ small we can write
\begin{eqnarray}
( V_{CKM})_{11}=\cos\theta_{12}=V'_{11}e^{i(\alpha_1-\beta_1)}=|V'_{11}|e^{i(\psi_1+\alpha_1-\beta_1)}\nonumber\\
( V_{CKM})_{12}=\sin\theta_{12}=V'_{12}e^{i(\alpha_1-\beta_2)}=|V'_{12}|e^{i(\phi_1+\alpha_1-\beta_2)}\nonumber\\
(V_{CKM})_{13}=\sin\theta_{13}e^{-i\delta_{CP}}=V'_{13}e^{i(\alpha_1-\beta_1)}=|V'_{13}|e^{i(\delta_1+\alpha_1-\beta_3)},
\label{ckm} 
\end{eqnarray} 
and more six relations. But these three are sufficient for our discussions. The phases associated to $V'$ elements, like $\psi_1$, $\phi_1$, and $\delta_1$ associated to $V'_{11}$, $V'_{12}$ and $V'_{13}$ respectively, are completely determinable in terms of $k$, $\phi$ using functional form of $\e$, $X$, $Y$, $Z$ and their associates phases. Now from the Eq.\ (\ref{ckm}) we obtain the following phase relations
\begin{eqnarray}
\psi_1+\alpha_1-\beta_1=0,\quad \phi_1+\alpha_1-\beta_2=0,\quad \delta_1+\alpha_1-\beta_3=-\delta_{CP}.
 \label{prl1} 
\end{eqnarray} 
Now the form of total mixing matrix is,
\begin{eqnarray}
V&=&V'V_P=U_{P1}^\dagger V_{CKM}U_{P2}V_P=U_{P1}^\dagger V_{CKM}\left(\begin{array}{ccc}
e^{i(\pi/2-\alpha+\beta_1)} & 0 & 0 \\
0 & e^{i(\pi/2+\beta_2)} & 0 \\
0  & 0 & e^{i(\pi/2-\gamma+\beta_3)}\end{array}\right) \nonumber\\
&&=\{e^{i(\pi/2-\gamma+\beta_3)}U_{P1}^\dagger\}V_{CKM}\left(\begin{array}{ccc}
e^{i(\beta_1-\beta_3+\gamma-\alpha)} & 0 & 0 \\
0 & e^{i(\gamma+\beta_2-\beta_3)} & 0 \\
0  & 0 & 1 \end{array}\right)
 \label{ffmv} 
\end{eqnarray} 
 This phase part in the parenthesis can be absorbed to charged lepton fields and the remaining part gives the leptonic mixing matrix of the form $V_{PMNS}=V_{CKM}\times {\rm diag}(e^{i\alpha_M},~e^{i\beta_M},~ 1)$, where the $\alpha_M$ and $\beta_M$ are the two Majorana phases of leptonic mixing matrix. From  Eq.\ (\ref{ffmv}) and using relations in Eq.\ (\ref{prl1}) and Eq.\ (\ref{phh}), we have the Majorana phases,
\begin{eqnarray}
\alpha_M=-\delta_{CP}-\frac{\Phi_2}{2}+\Psi \qquad \beta_M=-\delta_{CP}-\frac{\Phi_3}{2}+\Phi
 \label{majp2} 
\end{eqnarray} 
where $\Psi=\psi_1-\delta_1$ and $\Phi=\phi_1-\delta_1$ are known function of $k$ and $\phi$. In Eq.\ (\ref{majp2}) we have the relation of low energy and high energy phases. So, in our model we have correlation among the CP violating phases.
\section{Conclusion}
\label{sec:conlud}
We have shown that non-zero $U_{e3}$ is generated in a softly broken $A_4$ symmetric model through see-saw mechanism incorporating single parameter perturbation in $m_D$ in single element. First, we have studied all possible nine cases to explore the mixing angles considering all model parameters real. The extent of $\theta_{13}$ investigated, keeping the experimental values of present solar and atmospheric mixing angles. Among all nine possible texture of $m_D$ some of them generates non-zero $\theta_{13}$. Out of those non-zero  $\theta_{13}$ generating textures of $m_D$ we find  that breaking at '12' and '13' elements are encompassing the best values of $\theta_{12}$ and   $\theta_{23}$. However, the reach of $\theta_{13}$ in those cases are around $1^\circ$. Considering one of the parameter complex we extend our analysis with one of the most suitable texture of $m_D$ with breaking at '13' element. We have calculated mixing angles and neutrino mass squared differences in terms four model parameters ($m_0$, $\e$, $k$, $\phi$). We restrict model parameters utilising the well measured quantities $\Ds$, $\Da$,  $\theta_{12}$ and  $\theta_{23}$ and we have obtained $\theta_{13}$ (upto $2^\circ$) and large $J_{CP}$ ($\simeq 10^{-2}$)and $|(M_\nu)_{ee}|$ well below the present experimental upper bound. In addition to that a large $\delta_{CP}$ is also obtained. Further study on leptogenesis is also done and the present WMAP value of baryon asymmetry is obtained for a right handed neutrino mass scale $M_0\simeq 10^{13}$ GeV. In our model, we have seen that the phases responsible for the leptogenesis are correlated. We also find out the relations among low energy CP violating phases and the lepton asymmetry phases. Small $A_4$ symmetry breaking parameter $\e$, is sufficient to describe the all low energy neutrino data and high energy CP violation (leptogenesis). So, $A_4$ symmetry is an approximate symmetry.
\appendix
\section{}
Here we consider breaking of $A_4$ symmetry in all other entries 
of $m_D$. Case (i) is already discussed in the text.  
\vskip 0.1in
\noindent
(ii)Breaking at '11' element : In this case $m_D$ is given by
\begin{eqnarray}
 m_D = \frac{fv_u}{\surd 2}\pmatrix{1+\e&0&0\cr
                      0&1&0\cr
                      0&0&1}.
\label{bmd11}
\end{eqnarray} 
The mass eigenvalues are
\begin{eqnarray}
m_1 = -\frac{f^2v_u^2}{2(D+A)}\left(1+\frac{4\e}{3}\right)
\quad\quad m_2 = -\frac{f^2v^2_u}{2A}\left(1 + \frac{2\e}{3}\right)
\quad\quad m_3 = -\frac{f^2v^2_u}{2(D-A)}\nonumber\\
\end{eqnarray} 
and the three mixing angles come out as  
\begin{eqnarray}
\sin\theta_{12} = \frac{1}{\sqrt 3} +
\frac{2(2+k)}{3{\sqrt 3}k}\e
\quad\quad
\sin\theta_{23} = -\frac{1}{\sqrt 2}
\quad\quad 
\sin\theta_{13} = 0
\label{angel11}
\end{eqnarray} 
The solar and atmospheric mass differences and their ratio are
\begin{eqnarray}
&&\Ds = \frac{m_0^2}{3(1+k)^2}\left[
3k(k+2)+4\e(k^2+2k-1)\right]\nonumber\\
&&\Da = \frac{m_0^2}{3(1-k)^2}\left[
3k(2-k)-4\e(k-1)^2\right]\nonumber\\
&&R = \frac{\Ds}{\Da} 
  = \frac{(k -1)^2}{(k+1)^2}
    \frac{\left[3k(k+2)+4\e(k^2+2k-1)\right]}
    {\left[3k(2-k)-4\e(k-1)^2\right]}
\label{msd11} 
\end{eqnarray} 
The obtained expression for $\e$ in terms of model parameter $k$ and experimentally known $R$:
\begin{eqnarray}
\e = \frac{3k}{4(k-1)^2}
     \frac{\left[
    R(2-k)(k+1)^2-(k-1)^2(k+2)\right]}
    {\left[R(k+1)^2+k^2+2k-1\right]} 
\label{ep11} 
\end{eqnarray} 
iii) Breaking at '33' element :
In this case, the structure of 
$m_D$ is given by 
\begin{eqnarray}
 m_D = \frac{fv_u}{\surd 2}\pmatrix{1&0&0\cr
                      0&1&0\cr
                      0&0&1+\epsilon}.
\label{bmd33} 
\end{eqnarray} 
Mass eigenvalues  are 
\begin{eqnarray}
m_1 = -\frac{f^2v_u^2}{2(D+A)}\left(1+\frac{\e}{3}\right)
\quad\quad m_2 = -\frac{f^2v^2_u}{2A}\left(1 + \frac{2\e}{3}\right)
\quad\quad m_3 = -\frac{f^2v^2_u}{2(D-A)}\left(1+\e\right)\nonumber\\
\end{eqnarray}
and the angels are
\begin{eqnarray}
 \sin\theta_{12} = \frac{1}{\sqrt 3} - 
\frac{2+k}{3{\sqrt 3}k}\e
\quad 
\sin\theta_{23} = -\left[\frac{1}{\sqrt 2} - \e
\frac{k(4-k)}{6\sqrt 2(2-k)}\right]
\quad 
\sin\theta_{13} = 
-\frac{\e k(1-k)}{3{\sqrt 2}(2-k)}
\label{angel33} 
\end{eqnarray} 
The solar and atmospheric mass differences and their ratio are
\begin{eqnarray}
&&\Ds = \frac{m_0^2}{3(1+k)^2}\left[
3k(k+2)+2\e(2k^2+4k+1)\right]\nonumber\\
&&\Da = \frac{m_0^2}{3(1-k)^2}\left[
3k(2-k)-2\e(2k^2-4k-1)\right]\nonumber\\
&&R = \frac{\Ds}{\Da} 
  = \frac{(k -1)^2}{(k+1)^2}
    \frac{\left[3k(k+2)+2\e(2k^2+4k+1)\right]}
    {\left[3k(2-k)-2\e(2k^2-4k-1)\right]}
\label{msd33} 
\end{eqnarray} 
The obtained expression for $\e$ in terms of model parameter $k$ and experimentally known $R$:
\begin{eqnarray}
\e = \frac{3k}{2}
     \frac{\left[
    R(2-k)(k+1)^2-(k-1)^2(k+2)\right]}
    {\left[(k-1)^2(2k^2+4k+1)+
    R(k+1)^2(2k^2-4k-1)\right]}
\label{ep33} 
\end{eqnarray} 
iv) Breaking at '12' element :
In this case, the structure of 
$m_D$ is given by 
\begin{eqnarray}
m_D = \frac{fv_u}{\surd 2}\pmatrix{1&\epsilon&0\cr
                      0&1&0\cr
                      0&0&1} 
\label{bmd12} 
\end{eqnarray} 
Mass eigenvalues  are 
\begin{eqnarray}
m_1 = -\frac{f^2v_u^2}{2(D+A)}\left(1-\frac{2\e}{3}\right)
\quad\quad m_2 = -\frac{f^2v^2_u}{2A}\left(1 + \frac{2\e}{3}\right)
\quad\quad m_3 = -\frac{f^2v^2_u}{2(D-A)}\nonumber\\
\end{eqnarray} 
and the three mixing angles come out as 
\begin{eqnarray}
\sin\theta_{12} = \frac{1}{\sqrt 3} + 
\frac{2k+1}{3{\sqrt 3}k}\e
\quad
\sin\theta_{23} = -\left[\frac{1}{\sqrt 2} - \e
\frac{k(1-k)}{6\sqrt 2(2-k)}\right]
\quad
\sin\theta_{13} = 
-\frac{\e}{3\sqrt 2}\frac{k^2-k-3}{(k-2)}\nonumber\\
\label{angel12} 
\end{eqnarray} 
The solar and atmospheric mass differences and their ratio are
\begin{eqnarray}
&&\Ds = \frac{m_0^2}{3(1+k)^2}\left[
3k(k+2)+4\e(k^2+2k+2)\right]\nonumber\\
&&\Da = \frac{m_0^2}{3(1-k)^2}\left[
3k(2-k)-4\e(1-k)^2)\right]\nonumber\\
&&R = \frac{\Ds}{\Da} 
  = \frac{(k -1)^2}{(k+1)^2}
    \frac{\left[3k(k+2)+4\e(k^2+2k+2)\right]}
    {\left[3k(2-k)-4\e(1-k)^2\right]}
\label{msd12} 
\end{eqnarray} 
The obtained expression for $\e$ in terms of model parameter $k$ and experimentally known $R$:
\begin{eqnarray}
\e = \frac{3k}{4(1-k)^2}
     \frac{\left[
    R(2-k)(k+1)^2-(k-1)^2(k+2)\right]}
    {\left[(k^2+2k+2)+
    R(k+1)^2\right]}
\label{ep12} 
\end{eqnarray} 
v) Breaking at '13' element :
In this case, the structure of 
$m_D$ is given by 

\begin{eqnarray}
m_D = \frac{fv_u}{\surd 2}\pmatrix{1&0&\epsilon\cr
                      0&1&0\cr
                      0&0&1}
\label{bmd13} 
\end{eqnarray} 
The mass eigenvalues are
\begin{eqnarray}
m_1 = -\frac{f^2v_u^2}{2(D+A)}\left(1-\frac{2\e}{3}\right)
\quad\quad m_2 = -\frac{f^2v^2_u}{2A}\left(1 + \frac{2\e}{3}\right)
\quad\quad m_3 = -\frac{f^2v^2_u}{2(D-A)}\nonumber\\
\end{eqnarray} 
and the three mixing angles come out as 
\begin{eqnarray}
\sin\theta_{12} = \frac{1}{\sqrt 3} + 
\frac{2k+1}{3{\sqrt 3}k}\e
&&\quad \quad 
\sin\theta_{23} = -\left[\frac{1}{\sqrt 2} + \e
\frac{k(1-k)}{6\sqrt 2(2-k)}\right]\nonumber\\
\sin\theta_{13}&=& 
-\frac{\e}{3\sqrt 2}\frac{k^2-k-3}{(k-2)}
\label{angel13} 
\end{eqnarray} 
The solar and atmospheric mass differences and their ratio are
\begin{eqnarray}
&&\Ds = \frac{m_0^2}{3(1+k)^2}\left[
3k(k+2)+4\e(k^2+2k+2)\right]\nonumber\\
&&\Da = \frac{m_0^2}{3(1-k)^2}\left[
3k(2-k)-4\e(1-k)^2)\right]\nonumber\\
&&R = \frac{\Ds}{\Da} 
  = \frac{(k -1)^2}{(k+1)^2}
    \frac{\left[3k(k+2)+4\e(k^2+2k+2)\right]}
    {\left[3k(2-k)-4\e(1-k)^2\right]}
\label{msd13} 
\end{eqnarray} 
The obtained expression for $\e$ in terms of model parameter $k$ and experimentally known $R$:
\begin{eqnarray}
\e = \frac{3k}{4(1-k)^2}
     \frac{\left[
    R(2-k)(k+1)^2-(k-1)^2(k+2)\right]}
    {\left[(k^2+2k+2)+
    R(k+1)^2\right]}
\label{ep13} 
\end{eqnarray} 
vi) Breaking at '23' element :
In this case, the structure of 
$m_D$ is given by 
\begin{eqnarray}
m_D = \frac{fv_u}{\surd 2}\pmatrix{1&0&0\cr
                      0&1&\epsilon\cr
                      0&0&1}
\label{bmd23} 
\end{eqnarray} 
The mass eigenvalues are
\begin{eqnarray}
m_1 = -\frac{f^2v_u^2}{2(D+A)}\left(1+\frac{\e}{3}\right)
\quad\quad m_2 = -\frac{f^2v^2_u}{2A}\left(1 + \frac{2\e}{3}\right)
\quad\quad m_3 = -\frac{f^2v^2_u}{2(D-A)}\left(1-\e\right)\nonumber\\
\end{eqnarray} 
and the three mixing angles come out as 
\begin{eqnarray}
\sin\theta_{12} = \frac{1}{\sqrt 3} - 
\frac{\epsilon}{3{\sqrt 3}}\frac{2+k}{k}
\quad\quad 
\sin\theta_{23} = -\left[\frac{1}{\sqrt 2} - 
\frac{\epsilon}{2\sqrt 2}\right]
\quad\quad 
\sin\theta_{13} = 0
\label{angel23} 
\end{eqnarray} 
The solar and atmospheric mass differences and their ratio are
\begin{eqnarray}
&&\Ds = \frac{m_0^2}{3(1+k)^2}\left[
3k(k+2)+2\e(2k^2+4k+1)\right]\nonumber\\
&&\Da = \frac{m_0^2}{3(1-k)^2}\left[
3k(2-k)-2\e(2k^2-4k+5)\right]\nonumber\\
&&R = \frac{\Ds}{\Da} 
  = \frac{(k -1)^2}{(k+1)^2}
    \frac{\left[3k(k+2)+2\e(2k^2+4k+1)\right]}
    {\left[3k(2-k)-2\e(2k^2-4k+5)\right]}
\label{msd23} 
\end{eqnarray} 
The obtained expression for $\e$ in terms of model parameter $k$ and experimentally known $R$:
\begin{eqnarray}
\e = \frac{3k}{2}
     \frac{\left[
    R(2-k)(k+1)^2-(k-1)^2(k+2)\right]}
    {\left[R(1+k)^2(2k^2-4k+5)+
    (k-1)^2(2k^2+4k+1)\right]}
\label{ep23} 
\end{eqnarray} 
vii) Breaking at '21' element :
In this case, the structure of 
$m_D$ is given by 
\begin{eqnarray}
m_D = \frac{fv_u}{\surd 2}\pmatrix{1&0&0\cr
                      \e&1&0\cr
                      0&0&1}
\label{bmd21} 
\end{eqnarray} 
The mass eigenvalues are
\begin{eqnarray}
m_1 = -\frac{f^2v_u^2}{2(D+A)}\left(1-\frac{2\e}{3}\right)
\quad\quad m_2 = -\frac{f^2v^2_u}{2A}\left(1 + \frac{2\e}{3}\right)
\quad\quad m_3 = -\frac{f^2v^2_u}{2(D-A)}\nonumber\\
\end{eqnarray} 
and the three mixing angles come out as 
\begin{eqnarray}
\sin\theta_{12} = \frac{1}{\sqrt 3} + 
\frac{\epsilon}{3{\sqrt 3}}\frac{1-k}{k}
&&\quad \quad
\sin\theta_{23} = -\left[\frac{1}{\sqrt 2} + 
\frac{\epsilon}{6\sqrt 2}\frac{k(k-1)}{(2-k)}\right]\nonumber\\
\sin\theta_{13} &=& \frac{\epsilon}{3\sqrt 2}\frac{(3+k^2-4k)}
{(2-k)}
\label{angel21} 
\end{eqnarray} 
The solar and atmospheric mass differences and their ratio are
\begin{eqnarray}
&&\Ds = \frac{m_0^2}{3(1+k)^2}\left[
3k(k+2)+4\e(k^2+2k+2)\right]\nonumber\\
&&\Da = \frac{m_0^2}{3(1-k)^2}\left[
3k(2-k)-4\e(1-k)^2\right]\nonumber\\
&&R = \frac{\Ds}{\Da} 
  = \frac{(k -1)^2}{(k+1)^2}
    \frac{\left[3k(k+2)+4\e(k^2+2k+2)\right]}
    {\left[3k(2-k)-4\e(1-k)^2\right]}
\label{msd21} 
\end{eqnarray} 
The obtained expression for $\e$ in terms of model parameter $k$ and experimentally known $R$:
\begin{eqnarray}
\e = \frac{3k}{4(1-k)^2}
     \frac{\left[
    R(2-k)(k+1)^2-(k+2)(k-1)^2\right]}
    {\left[R(1+k)^2+k^2+2k+2
    \right]}
\label{ep21} 
\end{eqnarray} 
viii) Breaking at '31' element :
In this case, the structure of 
$m_D$ is given by 
\begin{eqnarray}
m_D = \frac{fv_u}{\surd 2}\pmatrix{1&0&0\cr
                      0&1&0\cr
                      \e&0&1}
\label{bmd31} 
\end{eqnarray} 
The mass eigenvalues are
\begin{eqnarray}
m_1 = -\frac{f^2v_u^2}{2(D+A)}\left(1-\frac{2\e}{3}\right)
\quad\quad m_2 = -\frac{f^2v^2_u}{2A}\left(1 + \frac{2\e}{3}\right)
\quad\quad m_3 = -\frac{f^2v^2_u}{2(D-A)}\nonumber\\
\end{eqnarray} 
and the three mixing angles come out as 
\begin{eqnarray}
\sin\theta_{12} = \frac{1}{\sqrt 3} + 
\frac{\epsilon}{3{\sqrt 3}}\frac{1-k}{k}
&&\quad \quad
\sin\theta_{23} = -\left[\frac{1}{\sqrt 2} + 
\frac{\epsilon}{6\sqrt 2}\frac{k(1-k)}{(2-k)}\right]\nonumber\\
\sin\theta_{13} &=& -\frac{\e}{3\sqrt{2}}
\frac{3+k^2-4k}{2-k}
\label{angel31} 
\end{eqnarray} 
The solar and atmospheric mass differences and their ratio are
\begin{eqnarray}
&&\Ds = \frac{m_0^2}{3(1+k)^2}\left[
3k(k+2)+4\e(k^2+2k+2)\right]\nonumber\\
&&\Da = \frac{m_0^2}{3(1-k)^2}\left[
3k(2-k)-4\e(1-k)^2\right]\nonumber\\
&&R = \frac{\Ds}{\Da}= \frac{(k -1)^2}{(k+1)^2}
    \frac{\left[3k(k+2)+4\e(k^2+2k+2)\right]}
    {\left[3k(2-k)-4\e(1-k)^2\right]}               
\label{msd31} 
\end{eqnarray} 
The obtained expression for $\e$ in terms of model parameter $k$ and experimentally known $R$:
\begin{eqnarray}
\e = \frac{3k}{4(1-k)^2}
     \frac{\left[
    R(2-k)(k+1)^2-(k-1)^2(k+2)\right]}
    {\left[R(1+k)^2+k^2+2k+2
    \right]}
\label{ep31} 
\end{eqnarray} 
ix) Breaking at '32' element :
In this case, the structure of 
$m_D$ is given by 
\begin{eqnarray}
m_D = \frac{fv_u}{\surd 2}\pmatrix{1&0&0\cr
                      0&1&0\cr
                      0&\e&1}
\label{bmd32} 
\end{eqnarray} 
The mass eigenvalues are
\begin{eqnarray}
m_1 = -\frac{f^2v_u^2}{2(D+A)}\left(1+\frac{\e}{3}\right)
\quad\quad m_2 = -\frac{f^2v^2_u}{2A}\left(1 + \frac{2\e}{3}\right)
\quad\quad m_3 = -\frac{f^2v^2_u}{2(D-A)}\left(1-\e\right)\nonumber\\
\end{eqnarray} 
and the three mixing angles come out as 
\begin{eqnarray}
\sin\theta_{12} = \frac{1}{\sqrt 3} - 
\frac{\epsilon}{3{\sqrt 3}}\frac{2+k}{k}
\quad 
\sin\theta_{23} = -\left[\frac{1}{\sqrt 2} + 
\frac{\epsilon}{2\sqrt 2}\right]
\quad 
\sin\theta_{13} = 0
\label{angel32} 
\end{eqnarray} 
The solar and atmospheric mass differences and their ratio are
\begin{eqnarray}
&&\Ds = \frac{m_0^2}{3(1+k)^2}\left[
3k(k+2)+2\e(2k^2+4k+1)\right]\nonumber\\
&&\Da = \frac{m_0^2}{3(1-k)^2}\left[
3k(2-k)-2\e(2k^2-4k+5)\right]\nonumber\\
&&R = \frac{\Ds}{\Da} 
  = \frac{(k -1)^2}{(k+1)^2}
    \frac{\left[3k(k+2)+2\e(2k^2+4k+1)\right]}
    {\left[3k(2-k)-2\e(2k^2-4k+5)\right]}
\label{msd32} 
\end{eqnarray} 
The obtained expression for $\e$ in terms of model parameter $k$ and experimentally known $R$:
\begin{eqnarray}
\e = \frac{3k}{2}
     \frac{\left[
    R(2-k)(k+1)^2-(k-1)^2(k+2)\right]}
    {\left[R(1+k)^2(2k^2-4k+5)+(1-k)^2
    (2k^2+4k+1)
    \right]}
\label{ep32} 
\end{eqnarray}



\begin{thebibliography}{99}
\bibitem{lelmlt}
S.~T.~Petcov,
Phys.\ Lett.\ B {\bf 110}, 245 (1982); for more recent studies  
see, e.g., 
R.~Barbieri \textit{et al.}, 
JHEP {\bf 9812}, 017 (1998); 
A.~S.~Joshipura and S.~D.~Rindani, 
Eur.\ Phys.\ J.\ C {\bf 14}, 85 (2000); 
R.~N.~Mohapatra, A.~Perez-Lorenzana and C.~A.~de Sousa Pires, 
Phys.\ Lett.\ B {\bf 474}, 355 (2000); 
Q.~Shafi and Z.~Tavartkiladze, 
Phys.\ Lett.\ B {\bf 482}, 145 (2000).
L.~Lavoura,
Phys.\ Rev.\ D {\bf 62}, 093011 (2000); 
W.~Grimus and L.~Lavoura,
Phys.\ Rev.\ D {\bf 62}, 093012 (2000); 
T.~Kitabayashi and M.~Yasue, 
Phys.\ Rev.\ D {\bf 63}, 095002 (2001); 
A.~Aranda, C.~D.~Carone and P.~Meade,
Phys.\ Rev.\ D {\bf 65}, 013011 (2002); 
K.~S.~Babu and R.~N.~Mohapatra, 
Phys.\ Lett.\ B {\bf 532}, 77 (2002); 
H.~J.~He, D.~A.~Dicus and J.~N.~Ng, 
Phys.\ Lett.\ B {\bf 536}, 83 (2002)
H.~S.~Goh, R.~N.~Mohapatra and S.~P.~Ng, 
Phys.\ Lett.\ B {\bf 542}, 116 (2002); 
G.~K.~Leontaris, J.~Rizos and A.~Psallidas, 
Phys.\ Lett.\ B {\bf 597}, 182 (2004). 
L.~Lavoura and W.~Grimus,
JHEP {\bf 0009}, 007 (2000); 
hep-ph/0410279.
S.~T.~Petcov and W.~Rodejohann,
  Phys.\ Rev.\ D {\bf 71}, 073002 (2005)
  [arXiv:hep-ph/0409135].

\bibitem{lmmlt}
Biswajit Adhikary, Phys.\ Rev.\ D{\bf 74}, 033002, (2006)  
[arXiv:hep-ph/0604009].
  S.~Choubey and W.~Rodejohann,
  Eur.\ Phys.\ J.\ C {\bf 40}, 259 (2005)
  [arXiv:hep-ph/0411190].
E.~Ma, D.~P.~Roy and S.~Roy,
Phys.\ Lett.\ B {\bf 525}, 101 (2002). 
  W.~Rodejohann and M.~A.~Schmidt,
  arXiv:hep-ph/0507300.
  E.~J.~Chun and K.~Turzynski, Phys.\ Rev.\ D{\bf 76}, 053008, (2007)  
[arXiv:hep-ph/0703070].

\bibitem{mutau}
  W.~Grimus and L.~Lavoura,
  JHEP {\bf 0107}, 045 (2001)
  [arXiv:hep-ph/0105212].
  W.~Grimus and L.~Lavoura,
  Acta Phys.\ Polon.\ B {\bf 32}, 3719 (2001)
  [arXiv:hep-ph/0110041].
  W.~Grimus, S.~Kaneko, L.~Lavoura, H.~Sawanaka and M.~Tanimoto,
  JHEP {\bf 0601}, 110 (2006)
  [arXiv:hep-ph/0510326] and reference there. 
  A.~Ghosal,
  Mod.\ Phys.\ Lett.\ A {\bf 19}, 2579 (2004).
  T.~Kitabayashi and M.~Yasue,
  Phys.\ Lett.\ B {\bf 524}, 308 (2002)
  [arXiv:hep-ph/0110303].
  T.~Kitabayashi and M.~Yasue,
  Phys.\ Rev.\ D {\bf 67}, 015006 (2003)
  [arXiv:hep-ph/0209294].
  I.~Aizawa, M.~Ishiguro, T.~Kitabayashi and M.~Yasue,
  Phys.\ Rev.\ D {\bf 70}, 015011 (2004)
  [arXiv:hep-ph/0405201].
W.~Grimus and L.~Lavoura,
  J.\ Phys.\ G {\bf 30}, 1073 (2004)
  [arXiv:hep-ph/0311362].
  R.~N.~Mohapatra and P.~B.~Pal,
  ``Massive neutrinos in physics and astrophysics. Second edition,''
  World Sci.\ Lect.\ Notes Phys.\  {\bf 60}, 1 (1998)
  [World Sci.\ Lect.\ Notes Phys.\  {\bf 72}, 1 (2004)].


\bibitem{tribi} P. F. Harrison, D. H. Perkins, and W. G. Scott, Phys. Lett.
{\bf B 530}, 167 (2002); P. F. Harrison and W. G. Scott, arXiv:
hep-ph/0402006.
\bibitem{Ma:2001dn}
  E.~Ma and G.~Rajasekaran,
  Phys.\ Rev.\  D {\bf 64}, 113012 (2001)
  [arXiv:hep-ph/0106291].

\bibitem{Altarelli:2005yx}
  G.~Altarelli and F.~Feruglio,
  Nucl.\ Phys.\  B {\bf 741}, 215 (2006)
  [arXiv:hep-ph/0512103].

\bibitem{Bazzocchi:2007na}
  F.~Bazzocchi, S.~Kaneko and S.~Morisi,
  arXiv:0707.3032 [hep-ph].
\bibitem{Yin:2007rv}
  F.~Yin,
  Phys.\ Rev.\  D {\bf 75}, 073010 (2007)
  [arXiv:0704.3827 [hep-ph]].

\bibitem{Sawanaka:2007ab}
  H.~Sawanaka,
  Int.\ J.\ Mod.\ Phys.\  E {\bf 16}, 1383 (2007)
  [arXiv:hep-ph/0703234].

\bibitem{Morisi:2007ft}
  S.~Morisi, M.~Picariello and E.~Torrente-Lujan,
  Phys.\ Rev.\  D {\bf 75}, 075015 (2007)
  [arXiv:hep-ph/0702034].

\bibitem{Koide:2007kw}
  Y.~Koide,
  arXiv:hep-ph/0701018.

\bibitem{Volkas:2006mk}
  R.~R.~Volkas,
  arXiv:hep-ph/0612296.

\bibitem{He:2006et}
  X.~G.~He,
  Nucl.\ Phys.\ Proc.\ Suppl.\  {\bf 168}, 350 (2007)
  [arXiv:hep-ph/0612080].

   \bibitem{Ma:2006vq}
  E.~Ma,
  Mod.\ Phys.\ Lett.\  A {\bf 22}, 101 (2007)
  [arXiv:hep-ph/0610342].

\bibitem{King:2006np}
  S.~F.~King and M.~Malinsky,
  Phys.\ Lett.\  B {\bf 645}, 351 (2007)
  [arXiv:hep-ph/0610250].

\bibitem{Lavoura:2006hb}
  L.~Lavoura and H.~Kuhbock,
  Mod.\ Phys.\ Lett.\  A {\bf 22}, 181 (2007)
  [arXiv:hep-ph/0610050].

\bibitem{Adhikary:2006jx}
  B.~Adhikary and A.~Ghosal,
  Phys.\ Rev.\  D {\bf 75}, 073020 (2007)
  [arXiv:hep-ph/0609193].

\bibitem{Ma:2006wm}
  E.~Ma,
  Mod.\ Phys.\ Lett.\  A {\bf 21}, 2931 (2006)
  [arXiv:hep-ph/0607190].

\bibitem{Ma:2006sk}
  E.~Ma, H.~Sawanaka and M.~Tanimoto,
  Phys.\ Lett.\  B {\bf 641}, 301 (2006)
  [arXiv:hep-ph/0606103].
  
\bibitem{Ma:2006dn}
  E.~Ma,
  Phys.\ Rev.\  D {\bf 73}, 057304 (2006).

\bibitem{Adhikary:2006wi}
  B.~Adhikary, B.~Brahmachari, A.~Ghosal, E.~Ma and M.~K.~Parida,
  Phys.\ Lett.\  B {\bf 638}, 345 (2006)
  [arXiv:hep-ph/0603059].

\bibitem{He:2006dk}
  X.~G.~He, Y.~Y.~Keum and R.~R.~Volkas,
  JHEP {\bf 0604}, 039 (2006)
  [arXiv:hep-ph/0601001].

\bibitem{Chen:2005jm}
  S.~L.~Chen, M.~Frigerio and E.~Ma,
  Nucl.\ Phys.\  B {\bf 724}, 423 (2005)
  [arXiv:hep-ph/0504181].

\bibitem{Krolikowski:2005ma}
  W.~Krolikowski,
  arXiv:hep-ph/0501008.

\bibitem{Ma:2004zv}
  E.~Ma,
  Phys.\ Rev.\  D {\bf 70}, 031901 (2004)
  [arXiv:hep-ph/0404199].

\bibitem{Hirsch:2003dr}
  M.~Hirsch, J.~C.~Romao, S.~Skadhauge, J.~W.~F.~Valle and A.~Villanova del Moral,
  Phys.\ Rev.\  D {\bf 69}, 093006 (2004)
  [arXiv:hep-ph/0312265].

\bibitem{Hirsch:2003xx}
  M.~Hirsch, J.~C.~Romao, S.~Skadhauge, J.~W.~F.~Valle and A.~Villanova del Moral,
  arXiv:hep-ph/0312244.

\bibitem{Ma:2003hu}
  E.~Ma,
  [arXiv:hep-ph/0311215].

\bibitem{Babu:2002dz}
  K.~S.~Babu, E.~Ma and J.~W.~F.~Valle,
  Phys.\ Lett.\  B {\bf 552}, 207 (2003)
  [arXiv:hep-ph/0206292].

\bibitem{ZEE}
  Xiao-Gang. He and A. Zee
  [arXiv:hep-ph/0607163].
\bibitem{Ma:2002yp}
  E.~Ma,
  Mod.\ Phys.\ Lett.\  A {\bf 17}, 627 (2002)
  [arXiv:hep-ph/0203238].
\bibitem{a4}
E. Ma, Phys. Rev. {\bf D 70}, 031901 (2004),
E. Ma, Phys. Rev. {\bf D 72}, 037301 (2005).
\bibitem{a4af}G. Altarelli and F. Feruglio, Nucl. Phys. {\bf B 720},
64 (2005); See also K. S. Babu and X.-G. He,
hep-ph/0507217.

\bibitem{Bazzocchi:2008rz}
  F.~Bazzocchi, S.~Morisi, M.~Picariello and E.~Torrente-Lujan,
  arXiv:0802.1693 [hep-ph].
\bibitem{Altarelli:2008bg}
  G.~Altarelli, F.~Feruglio and C.~Hagedorn,
  arXiv:0802.0090 [hep-ph].
\bibitem{Brahmachari:2008fn}
  B.~Brahmachari, S.~Choubey and M.~Mitra,
  arXiv:0801.3554 [hep-ph].

\bibitem{Honda:2008rs}
  M.~Honda and M.~Tanimoto,
  arXiv:0801.0181 [hep-ph].

\bibitem{Lavoura:2007dw}
  L.~Lavoura and H.~Kuhbock,
  arXiv:0711.0670 [hep-ph].

\bibitem{EspositoFarese:2007sw}
  G.~Esposito-Farese,
  arXiv:0711.0332 [gr-qc].

\bibitem{Ma:2007hg}
  E.~Ma,
  arXiv:0710.3851 [hep-ph].

\bibitem{Baunack:2007zz}
  S.~Baunack,
  Eur.\ Phys.\ J.\  A {\bf 32}, 457 (2007).

\bibitem{Bazzocchi:2007au}
  F.~Bazzocchi, S.~Morisi and M.~Picariello,
  Phys.\ Lett.\  B {\bf 659}, 628 (2008)
  [arXiv:0710.2928 [hep-ph]].
\bibitem{Grimus:2008tm}
  W.~Grimus and H.~Kuhbock,
  arXiv:0710.1585 [hep-ph].
\bibitem{Plentinger:2008up}
  F.~Plentinger, G.~Seidl and W.~Winter,
  JHEP {\bf 0804}, 077 (2008)
  [arXiv:0802.1718 [hep-ph]].
\bibitem{branco1}G. C. Branco, T. Morozumi, B. M. Nobre, M. N. Rebelo,
hep-ph/0107164, hep-ph/0202036, R. N. Mohapatra and W. Rodejohann,
hep-ph/0507312.


\bibitem{Pilaftsis:2003gt}
  A.~Pilaftsis and T.~E.~J.~Underwood,
  %
  Nucl.\ Phys.\ B {\bf 692}, 303 (2004)
  [arXiv:hep-ph/0309342].
 
\bibitem{Fukugita:1986hr}
  M.~Fukugita and T.~Yanagida,
  %
  Phys.\ Lett.\ B {\bf 174}, 45 (1986).

\bibitem{leptogen}
  M.~A.~Luty,
  %
  Phys.\ Rev.\ D {\bf 45}, 455 (1992).
  M.~Flanz, E.~A.~Paschos and U.~Sarkar,
  %
  Phys.\ Lett.\ B {\bf 345}, 248 (1995)
  [Erratum-ibid.\ B {\bf 382}, 447 (1996)]
  [arXiv:hep-ph/9411366].
  M.~Plumacher,
  %
  Z.\ Phys.\ C {\bf 74}, 549 (1997)
  [arXiv:hep-ph/9604229].
  L.~Covi, E.~Roulet and F.~Vissani,
  %
  Phys.\ Lett.\ B {\bf 384}, 169 (1996)
  [arXiv:hep-ph/9605319].
  W.~Buchmuller and M.~Plumacher,
  %
  Phys.\ Lett.\ B {\bf 431}, 354 (1998)
  [arXiv:hep-ph/9710460].

\bibitem{Pilaftsis1}
  A.~Pilaftsis,
  %
  Int.\ J.\ Mod.\ Phys.\ A {\bf 14}, 1811 (1999)
  [arXiv:hep-ph/9812256].
  W.~Buchmuller and M.~Plumacher,
  %
  Int.\ J.\ Mod.\ Phys.\ A {\bf 15}, 5047 (2000)
  [arXiv:hep-ph/0007176].
  E.~A.~Paschos,
  %
  Pramana {\bf 62}, 359 (2004)
  [arXiv:hep-ph/0308261].

\bibitem{Nielsen:2001fy}
  H.~B.~Nielsen and Y.~Takanishi,
  %
  Phys.\ Lett.\ B {\bf 507}, 241 (2001)
  [arXiv:hep-ph/0101307].

\bibitem{Roos:1994fz}
  M.~Roos 1994,
  ``Introduction to cosmology,'' (John Wiley \& Sons)


\bibitem{Harvey:1990qw}
  J.~A.~Harvey and M.~S.~Turner,
  Phys.\ Rev.\ D {\bf 42}, 3344 (1990).

\bibitem{Barger}
  V.~Barger, D.~A.~Dicus, H.~J.~He and T.~j.~Li,
  Phys.\ Lett.\ B {\bf 583}, 173 (2004)
  [arXiv:hep-ph/0310278].
  W.~Buchmuller, P.~Di Bari and M.~Plumacher,
  Nucl.\ Phys.\ B {\bf 643}, 367 (2002)
  [arXiv:hep-ph/0205349].
%
\bibitem{kolb}
E. W. Kolb and M. S. Turner 1990, ``The Early Universe''
(Addison-Wesley)

\bibitem{Giudice:2003jh}
  G.~F.~Giudice, A.~Notari, M.~Raidal, A.~Riotto and A.~Strumia,
  Nucl.\ Phys.\ B {\bf 685}, 89 (2004)
  [arXiv:hep-ph/0310123].

\end{thebibliography}
\end{document}